\pdfoutput=1
\documentclass[11pt]{article}
\usepackage[letterpaper, left=1in, right=1in, top=1in,bottom=1in]{geometry}

\usepackage[linesnumbered, ruled, vlined]{algorithm2e}

\SetAlFnt{\small}
\SetAlCapFnt{\small}
\SetAlCapNameFnt{\small}
\SetAlCapHSkip{0pt}
\IncMargin{-\parindent}

\usepackage{verbatim}
\usepackage{natbib}

\makeatother
\usepackage{amsmath,amsfonts}
\usepackage{amsmath,amstext,rotating}
\usepackage{amsfonts,amssymb,graphics,xspace,endnotes}
\usepackage{fancyhdr}
\usepackage{epsfig}
\usepackage{color}            

\usepackage[suppress]{color-edits} 
\addauthor{et}{blue}
\addauthor{etc}{red}
\addauthor{dn}{green}
\addauthor{pj}{magenta}

\usepackage{geometry}
\usepackage[colorlinks,linkcolor=black,citecolor=black,urlcolor=black,pdfstartview=FitH,bookmarks=false]{hyperref}
\newcommand{\ba}{\begin{array}}
\newcommand{\ea}{\end{array}}
\newcommand{\bs}{\begin{align}\begin{split}\nonumber}
\newcommand{\bsnumber}{\begin{align}\begin{split}}
\newcommand{\es}{\end{split}\end{align}}

\newtheorem{theorem}{THEOREM}
\newtheorem{definition}{DEFINITION}

\newcommand{\NR}{{\mathcal NR}}
\newcommand{\R}{{\mathbb R}}

\begin{document}
\title{Learning and Trust in Auction Markets}



\author{%
Pooya Jalaly \thanks{Email: \texttt{jalaly@cs.cornell.edu}. Work supported in part by NSF grant CCF-1563714, ONR grant N00014-08-1-0031, and a Google Research Grant. }
\and
Denis Nekipelov \thanks{Department of Economics,  University of Virginia, Monroe Hall, Charlottesville, VA 22904, USA. Email: \texttt{ denis@virginia.edu}. Work supported in part by NSF grant CCF-1563714 , and a Google Research Grant. }
	\and
	\'{E}va Tardos\thanks{Department of Computer Science, Cornell University, Gates Hall, Ithaca, NY 14853, USA, Email: \texttt{eva@cs.cornell.edu}. Work supported in part by NSF grant CCF-1563714, ONR grant N00014-08-1-0031, and a Google Research Grant.}
}
\maketitle
\begin{abstract}
Auction theory analyses market designs by assuming all players are fully rational.
In this paper we study behavior of bidders in an experimental launch of a new advertising
auction platform by Zillow, as Zillow switched from negotiated contracts to using auctions
in several geographically isolated markets. A unique feature of this experiment is that the
bidders in this market are local real estate agents that bid in the auctions on their own behalf,
not using third-party intermediaries to facilitate the bidding. To help bidders,  Zillow also provided a recommendation tool that suggested the bid for each bidder.

Our main focus in this paper is on the decisions of bidders whether or not to adopt the platform-provided bid recommendation. We observe that a significant proportion of bidders do not use the recommended bid. Using the bid history of the agents we infer their value, and compare the agents' regret with their actual bidding history with results they would have obtained consistently following the recommendation. We find that for half of the agents not following the recommendation, the increased effort of experimenting with alternate bids results in increased regret, i.e., they get decreased net value out of the system. The proportion of agents not following the recommendation slowly declines as markets mature, but it remains large in most markets that we observe. We argue that the main reason for this phenomenon is the lack of trust that the bidders have in the platform-provided tool.

Our work provides an empirical insight into possible design choices for auction-based online advertising platforms. While search advertising platforms (such as Google or Bing) allow bidders to submit bids on their own and there is an established market of third-party intermediaries that help bidders to bid over time, many display advertising platforms (such as Facebook) optimize bids on bidders' behalf and eliminate the need for the bidders to bid on their own or use intermediaries. Our empirical analysis shows that the latter approach is preferred for markets where bidders are individuals, who don't have access to third party  tools, and who may question the fairness of platform-provided suggestions.
\end{abstract}
\newpage

\section{Introduction}
Auction theory analyses market design by assuming all players behave fully rationally, and the outcome is a (Bayes) Nash equilibrium of the game. Some recent work, such as \cite{NST:2015}, suggests to replace this assumption for repeated games (such as ad-auctions) with modeling the players as learners, assuming they use a form of no-regret learning in repeated games to find the best strategy to play. No-regret learning can be implemented with less available information, but the assumption is still modeling agents with a strong form of rationality, and using high level of data analytics. This assumption is well justified in auctions where bidders use strong tools for data analytics, or have a market place of third-party intermediaries to facilitate the bidding, and bidders invest enough in the market to pay for the analytics. Bidders in these market places use algorithmic bidding tools, and such tools do optimize rationally, and hence are much less subject to human biases.

In this paper we study bids in an experiment with auction where bidders are humans, each with relatively small investment, and were not using algorithmic tools. In such auctions, the reality may challenge the above assumptions. The actions of human bidders, not assisted by strong analytical tools, may be affected by issues not considered in classical auction theory: the bidders will lack information and lack the attention needed to make rational decisions, and may also be effected by behavioral biases that are not accounted for in the standard theory.

Our data comes from an experimental launch of a new advertising auction platform by Zillow. Zillow.com is the largest residential real estate search platform in the United States used by 140 million of people each month according to the company's statistic \cite{zillow-statistics}. Viewers are looking to buy or sell houses, want to see available properties, typical prices, and learn about market characteristic. The platform is monetized by showing ads of real estate agents offering their services. Historically, Zillow used negotiated contracts with real-estate agents for placing ads on the platform. 
In the experiment we study, several geographically isolated markets were switched from negotiated contracts to auction based pricing and allocation. 
The auction design used was a form of generalized second price, very similar to what is used in many other markets except that agents were paying for impressions (and not only for clicks). A unique feature of this experiment is that the bidders in this market are local real estate agents that bid in the auctions on their own behalf. This is unlike many existing online marketplaces where bidders use third-party intermediaries to facilitate the bidding. Along with the new auction platform, Zillow provided the bidders the recommendation tool that suggested the bid for each bidder based on the inputs of this bidder's target parameters (e.g. \etedit{impression volume,} 
budget, and competing bids of other bidders).

The main focus of our paper is understanding the bidder's decision whether or not to adopt the platform-provided bid recommendation. Bidders were required to log into the system if they wanted to change their bid, and once they logged in, the system offered a suggested bid: the recommended bid for maximizing the obtained impression volume for the bidders' budget. Our main conclusion is that the bidders lack trust in the recommendation, and both bidders and the platform would have been better off if the system didn't offer bidders the opportunity to avoid the recommended bid.

Our main metric for the analysis of the bidders' bid sequences is the average regret measuring the difference between the average utility that was achieved by the bid sequence and the utility from the best fixed bid in the hindsight. A fixed bid is not actually optimal in this environment, as bidding differently on different days of the week would have been beneficial for the bidders. However, regret for a fixed bid with hindsight seems to be the most fair comparison, as the recommendation tool was essentially making fix bid recommendations (a limitations of its design), and the bidder's behavior seems to be well approximated with looking for a good fix bids: they didn't update bids frequently enough to take advantage of the opportunities varying with the days of the week.

We observe that a large proportion of bidders does not use the recommended bid to make bid changes
immediately following the introduction to the new market.
Our main finding is that the observed bid sequences that deviated from recommended bids didn't typically result in smaller average regret than the recommended bid. In other words, even though many bidders attempted to adjust bids on their own, and they had the opportunity to gain over the recommended bid in terms of overall value (by bidding differently on weekdays and weekends), many bidders would have been better off by always using recommended bid. The number of bidders who outperformed the bid recommendation in our study is about the same as the number of those who did worse. The proportion of bidders following the recommendation slowly increases as markets mature, but it remains large in most markets that we observe. We argue that the main reason for this phenomenon is the lack of trust that the bidders have in the platform-provided tool.

An important challenge in understanding the data is the uncertainty in the bidders values for each impression. Most bidders in this market are limited by small budgets, and as a result their bid, even if interpreted as a fully rational learning behavior, may not have enough information to infer the value of the bidders. We use the the learning based inference of \cite{NST:2015} to infer the agent's values based on their bidding behavior, and then compare this regret, to the regret on this inferred value, had the follow the platform's bid recommendation. Note that the regret inferred for the bidders behavior is a lower bound on the actual regret: the value we infer for the player is the value that would give them the smallest possible regret. Our results show that under this value, they would have less regret had they adopted the platform recommendation.

Our work provides an empirical insight into possible design choices for auction-based online advertising platforms. Search advertising platforms (such as Google or Bing) allow bidders to submit bids on their own and there is an established market of third-party intermediaries that help bidders to bid over time. This market design allows for more complex bidding functions, for example allowing agents to express added value for subsets of the impression opportunities via multiplicative bid-adjustments (e.g., based on the age of the viewer). In contrast, many display advertising platforms (such as Facebook) use a simpler bidding language, and optimize bids on bidders' behalf based solely on their budgets. This eliminates the need for the bidders to bid on their own or use intermediaries. Our empirical analysis shows that, despite its more limited expressibility, the latter approach may be preferred for markets where bidders are individuals who don't have access to third party tools, and who may question the fairness of platform-provided suggestions.

\paragraph{\textbf{Related Work}}
Number of papers in recent years focus on estimating bidder's value in online advertising auctions. One the earliest papers in this area is \cite{AtheyNekipelov}, who study bidder values in Bing's GSP auction for search ads. They use the equilibrium characterization of GSP, and find that the bidders utility functions are smooth and strongly convex as the function of their bids. This ensures that if bids are at equilibrium, bidder valuations are uniquely identifiable based on the bid. In dynamic or new markets where interaction is repeated, the value of each individual interaction is small, and bidders are not (yet) knowledgeable about the system, it is better to model players as learners. \cite{NST:2015} suggest this assumption for studying bidders in Bing's market for search ads, and shows how to infer values based on bidding behaviour under this weaker assumption on the outcome. To evaluate the effectiveness of the bid-recommendation tool for the bidders, we need to estimate their value for impressions. We do this using the methodology developed in \cite{NST:2015}, making the assumption that agents are low-regret learners.

In a recent paper \cite{NisanN16} the authors report on a human subject experiment on the reliability of regret based inference. In their experiment, human subjects participated in bidding games (including the GSP format). The paper asks the question if human behavior can be modeled as no-regret learning, and to what extent the inference based
on the low regret assumption can be used to recover the bidders value from their bidding behavior. Their finding
are mixed. They find the players whose value is high behave rationally, experiment with the best bidding behaviour,  achieve very low regret, and inference based on this assumption accurately recovers their value. The finding for players with low types is less positive. Some participants in the experiments were given values so low, that rational behavior would have them drop out of the auction (or bid so low they are guaranteed to lose). Such low value players were frustrated by the game, and behaved rather irrationally at times. It is interesting to think about the contrast between the participants in the Nisan-Noti laboratory experiment and the agents in the Zillow field experiment.
The players in the Nisan-Noti experiment were paid to participate (even if frustrated), while in contrast participation in Zillow's ad-auctions is optional, and for typical real estate agents Zillow may not be the main channel through which they get the ``client leads''. Frustrated agents can  drop out, and in fact, there were many short lived agents in our data. We focus our analysis on agents that stay in the system for an extended period of time. In addition, we note that \cite{NisanN16} as well as \cite{NST:2015} identify the value with smallest regret error relative to the value. This method favors larger values, that make the relative error smaller. Using the value with smallest absolute error would have made the identification mode successful even for bidders with relatively smaller values. This is the method we will use in this paper.

A distinctive feature of Zillow's field experiment was that the bidders were provided the
bid recommendation tool. Such tools are not unique to Zillow and are
routine in search advertising on Google and Bing such as in \cite{google-tool}. 
\cite{LinkedIn16} report experiments with adding bid-recommendations at LinkedIn, where they find that the advertisers and the publisher both benefit having recommendations. On those platforms
there is also a set of third-party tools (not provided by platforms) that facilitate bidding.
However, on Zillow the bidders were
faced with the choice between trusting the recommendation provided by Zillow's tool
or learning on their own. Our work thus bridges the gap betwen the literature on
empirical analysis of algorithmic learning in games and the literature on recommender
systems without trust (e.g. see \cite{ricci:2010} for a survey of the latter).

\section{Auction design}
\begin{figure}[!ht]
\begin{center}
\includegraphics[width=.5\textwidth]{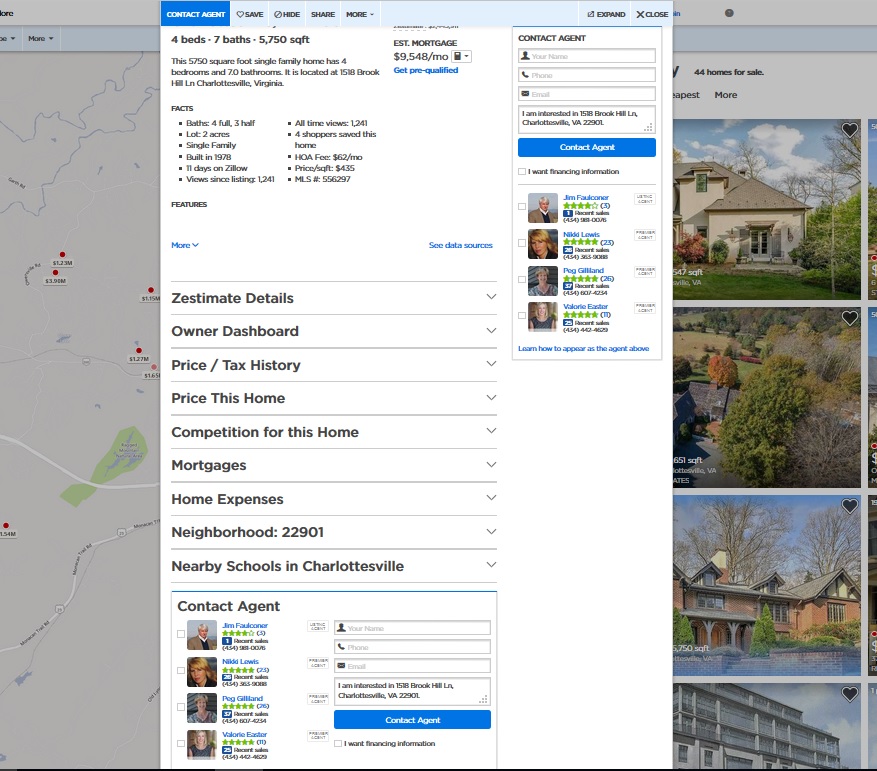}
\caption{Sample property search result on Zillow.com with highlighted premiere agents}\label{fig1}
\end{center}
\end{figure}

\paragraph{\textbf{Background.}}
Zillow.com is the largest residential real estate search platform in the United States. Like all of the big search platforms, it is ``consumer-facing": it offers consumers free interactive information about the current real estate listings, historical data on real estate sales, real estate valuation for the properties that are not currently for sale, the background local demographic information
that includes average incomes, age and education of residents as well as the measures of qualities of local schools. Similar to other Internet based services, 
 Zillow's business is based on
monetization of consumer page views by selling advertisement opportunities.
Whenever a consumer clicks on particular
property from the list of the search results, the page that opens gives details on the property.
In addition, on the right side and at the bottom of the page, a list
of real estate agents are shown. A sample page is shown on Figure \ref{fig1}.
The first agent on the list is always a listing agent for the property that that consumer
has clicked on (if the property is for sale). The rest of the agents listed (highlighted as ``premier agents'') are real estate agents
advertising their services to consumers viewing the listed property page.
At the time period when we collected our data, only three premier agents were shown
per page and the list of premier agents was identical on the side and at the bottom of the page.

Premier agents buy impressions from specific zip codes that they selected. Once the system identifies the agent's eligible for a given impressions, the agents shown on the page in random order.
As a result, in expectation all impressions in the given zip code have the same value (in contrast with Google where bidders are ordered by their bid, and higher positions are viewed as better).
Historically, as on many other consumer platforms on the Internet, these impressions were sold through negotiated contracts between Zillow and the real estate agents.

In order to improve fairness and efficiency of the market, in the period
between 2012 and 2015 Zillow engaged in a series of long-term experiments in which for a set of select
geographically distinct zip codes across the United States, the negotiated contract system for selling impressions was replaced with auction-based system.
The goal of these large scale experiments was twofold. On the one hand, the platform wanted to study the
revenue impact of switching from rigid system of long-term fixed price contracts to a dynamic auction
system that allows the impression prices to change in real time. On the other hand, they wanted
to use the bids to understand the discrepancy between the negotiated prices and the agent's values for impression.

The auction format used for the experiments was generalized second prize (GSP) auction for the slots of available ad positions.
Recall that the ad delivery system randomizes the order of the agents and so keeps all impressions the same expected quality for each agent.
Maybe the most natural mechanism would be to simply select the three highest bids, and price them at the uniform price of the 4th bid. However, \cite{chawla:16} shows that the use of uniform price mechanism may limit the quality of inference of bidder values from the bids, and a form of discriminative price mechanism, such as the GSP, allows for better inference of values. To distinguish the agents by their bid, for the agents with a bit lower bids, Zillow's mechanism decreased the probability of the agent's ad getting places on the page.
Below we outline the structure of the implemented mechanism and the structure of static best responses of bidders.

\paragraph{\textbf{The mechanism.}}
\label{Sec:AuctionBestResponses}
The mechanism implemented in the large scale experiments run by Zillow can be characterized as
\begin{itemize}
\setlength{\itemsep}{0pt}\setlength{\parsep}{0pt}\setlength{\parskip}{0pt}
\item real-time, with ads places in real time as opportunities arise
\item weighted, higher bids have a higher chance of being shown,
\item agents are paying per impression, unlike the per-click payment used for search ads
\item generalized second price auction
\item with reserve prices and budget smoothing.
\end{itemize}
In this mechanism each bidder $i$ submits her bid $b_i$ and (typically per month)
budget, though agents are allowed to submit budgets for shorter periods, and some do. The auction platform takes fixed position ``weights'' $\gamma_j$. These weights are used by the platform to induce the dependence
of the impression allocation on the rank of each bidder's bid: weights sum to 1, and the agents with $j$th highest bid is shown with probability $ 3\gamma_j$ on a page, so 3 agents are shown on each page. The weights used in the system are $0.33, 0.28, 0.22, 0.17$, so only the highest 4 bids have a chance of being shown.

Real-estate agents typically have relatively small budgets, so the system needs to implement a form of ''budget-smoothing'' or pacing to have the agents participate in auctions evenly across the time interval. For each bidder $j$, the system determines a budget-smoothing probability $\pi_j$, that in expectation will ensure that the agents don't overspend their budget.

The mechanism then implements the GSP {\it taking into account bidders' budgets.}
In each impression opportunity this is done as follows:
\begin{enumerate}
\setlength{\itemsep}{0pt}\setlength{\parsep}{0pt}\setlength{\parskip}{0pt}
\item The advertiser database is queried for all bidders eligible to be shown in a given impression opportunity, that is, advertisers bidding on the ZIP code of the property
\item For the set of eligible agents, the system determines the filtering probabilities for budget smoothing. To do this, the system needs to estimate the expected spent of the agent given her bid $b_i$. This turns out to be a fixed point computation, as each expected spend depends also on the filtering probabilities of other agents.
\item the remaining bidders are ranked by the order of their bids
\item Three of the top four remaining bidders are displayed, so that the probability that the ad of bidder ranked $j$ is shown is $3\gamma_j$
\item If the bidder ranked $j$ is shows, she pays the bid of the bidder ranked $j+1$ (or the reserve price) for the impression.
\end{enumerate}
To avoid having to deal with ties, Zillow effectively implemented a priority order via assigning each agent  ''quality score" very close to 1, to determine the order of agents with identical bids.

We describe the details of budget smoothing, and analyze the properties of the auction mechanism from the {\it expected impression} perspective.
Although, this gives a simplified view of the system (e.g. avoiding
dynamics and the fluctuation of the impression volume), that allows to discuss
the incentives in the auction mechanism in the most crystallized
way. If $NP$ is the total number of
page views in thousands over the time period of the bidder $i$ and $\bar{B}_i$ is the total budget of the bidder, then the {\it per thousand impression opportunity  budget}
can be expressed as
$
B_i=3\,\bar{B}_i\, \big/\, NP.
$
This reflects the fact that on each
page there are 3 ad impression opportunities for 3 available slots and
each bidder can only appear in one of the three slots. We further assume
that the bidders are risk-neutral and have quasi-linear utility, so their utility is characterized
by {\it values per thousand impressions} $v_i$.

In the per impression context the budget smoothing process
can be characterized by a probability $\pi_i$ that determines
eligibility of bidder $i$ for an auction. {\it Conditional on not being budget smoothed},
bidder $i$ participates in a generalized second price auction. Since all
bidders may be budget smoothed,
the group of auction participants becomes random. As a result,
the auction outcomes for bidder $i$ (her rank $j$, and so the probability $\gamma_j$ that her ad is being shown, as well as her price per impression) are random. Taking expectation over the
budget smoothing of other agents, we can construct the expected auction outcomes:
$\mbox{eCPM}_i(b_i)$ is the expected cost per a thousand impression opportunities. Note that this expectation is also a function of the other bids $b$ and the budget smoothing probabilities $\pi$ of opponent bidders, a dependence that we will make explicit in notation when useful. Similarly, the probability of 
the impression being shown is also a random variable which we denote by $eQ_i(b_i)$. We note that
both these objects are defined via the conditional expectation, i.e. they
determine the spend and the impression probability conditional on
bidder $i$ not being budget smoothed. The impression eligibility probability is determined by the balanced budget condition:
\begin{equation}\label{eq:smoothing}
\pi_i=\min\left\{1,\;\frac{B_i}{\mbox{eCPM}_i(b_i)}\right\}.
\end{equation}
Note that if the expected per impression spent does not
exceed the per impression budget, then such a bidder should
not ever be budget smoothed.

\paragraph{\textbf{Best Responses}.}
Now we can characterize the structure of the bidder's utility
and the best response bid. When $\mbox{eCPM}_i(b_i)<B_i$,
the expected utility per impression is determined solely by the expected auction
outcomes, i.e.
$$
u_i(v_i,B_i,b_i)=eQ_i(b_i)\,v_i-\mbox{eCPM}_i(b_i).
$$

Classical analysis of such an economic system would assume that the outcome is a Nash equilibrium of the game, where each
bidder maximizes her utility by setting the bid. Due to lack of space we will skip here the details of the resulting equilibrium analysis.
We note that identifying the right bid can be challenging for the bidders, who are real-estate agents, and often don't have the data or the analytic tools to do a good job optimizing their bid. To help the advertisers, the platform  provides a bid recommendation, suggesting the bid that maximizes the expected number of impression the agent can achieve based on her budget.

\paragraph{\textbf{Budget smoothing (Pacing).}}
\label{Sec:BudgetSmoothing}
Budget smoothing is one of the most technically challenging
components of the implemented experimental mechanism.
For large advertisers on platforms like Google, budgets typically play a
minor role essentially working as ``insurance'' from surges
in spent generated by idiosyncratic events. In contrast, advertisers on Zillow are real-estate agents, and typically have small monthly budgets
relative to per impression cost. In particular, our data shows that virtually all bidders are budget smoothed over certain periods.

In these settings a carefully constructed system for budget
smoothing is essential.
Take the vector of current eligible bids and budgets for bidders $i=1,\ldots,I$. The idea for recovering the filtering probabilities will be to solve for a set of probabilities $\pi_1,\ldots,\pi_I$ such that (\ref{eq:smoothing}) is satisfied for each $i$.
The main ingredient in computing $\pi_i$ is the expected cost per opportunity
$\mbox{eCPM}_i(b_i)$ with expectation taken  with respect to the distribution of $\pi_1,\ldots,\pi_I$ of other bidders, and $eQ_i(b_i)$ is the probability of being shown conditional on not being filtered out.
In Appendix \ref{smoothing:appendix} we give the algorithmic description of the computation of $\pi_i$'s.

\section{Market environment}


\paragraph{\textbf{Data description.}}
In the period between 2014 and 2016\footnote{We withhold the exact start and end date of the experiments
for confidentiality purposes} Zillow has run a series of large-scale experiments
where the mechanism for selling ad impressions was switched from negotiated contracts
to auctions. 
During this period,
Zillow defined markets as zip codes. The real estate agents were not allowed \pjedit{to} use targeting
within the zip code (i.e. by advertising only on the pages of specific real estate listings)
or buy ``packages'' of impressions across multiple zip codes. In fact, a vast majority of real estate agents
that we observe in the data only compete for a single zip code.

The experiments were rolled out in a large number of clearly isolated markets with zip codes
coming from either separate states or sufficiently far from each other within the state.
In order to facilitate this experimental mechanism rollout, Zillow has engaged in
a significant marketing and training effort to ensure that real estate agents in
the experimental markets 
understand the structure of the auction and to help agents learn how to bid well in the auction,
akin the set of tutorials provided by Google for its advertisers.

Our data comes from 57 experimental markets from Zillow. These markets are close
to the entirety of markets that were switched to auction-based prices and allocations.
We dropped a few markets from the data that
either did not have reliable data
due to possible malfunction of the implementation of the auction mechanism, or
the data span was too short to produce reliable results.
For data confidentiality purposes all dollar-valued variables, such as prices and
budgets, in our data were re-scaled and do not reflect the actual amounts.

Our structural analysis in this paper will be concentrated on the much smaller
set of 6 very active markets. Our goal in selecting these markets was to (i) ensure that those
markets are sufficiently geographically separated,
yet have the typical statistical
properties of all markets, such as impression prices, in all characteristics
except the activity of agents; (ii) have sufficient number of observations
of bid changes for different bidders. To understand the behavior of bidders in
these auctions, we need to infer their values. Agents that are not active on the
platform, do not provide enough data to reasonably estimate their value.
As it will become clear in Section 4, the
second part is crucial for us to be able to produce reliable evaluation of
payoffs and bidding strategies of the agents.
To select the 6 markets, we first filter the markets where the number of participating agents is 15 or less, which gets down the number of regions from 57 to 12. The average frequency of bid changes per day in these regions was 0.43. The 6 markets we use for our structural estimation, are the markets with above average number of bid changes.

{\small
\begin{center}
\begin{table}[!ht]
\begin{center}
\begin{tabular}{|l|l|l|l|l|l|l|l|l|}
\hline
Variable & \multicolumn{4}{|c|}{Selected Regions} & \multicolumn{4}{|c|}{All Regions} \\
\cline{2-9}
            & Mean & STD & 25\% & 75\%  & Mean & STD & 25\% & 75\%\\ \hline
Number of agents & 19.33 & 2.29 & 18.0 & 20.75 & 10.74 & 5.32 & 6.0 & 15.0 \\ \hline
Bids & 23.94 & 14.14 & 17.3 & 19.31 & 18.79 & 9.71 & 14.06 & 23.84 \\ \hline
Budgets (daily) & 8.92 & 3.0 & 6.31 & 11.71 & 9.22 & 4.96 & 5.9 & 12.44 \\ \hline
Active duration & 85.97 & 10.38 & 78.03 & 91.5 & 96.04 & 20.74 & 86.53 & 107.33 \\ \hline
Reserve price & 11.65 & 7.03 & 7.99 & 10.74 & 13.39 & 9.55 & 6.0 & 16.93 \\ \hline
Bid changes & 0.73 & 0.26 & 0.54 & 0.85 & 0.22 & 0.28 & 0.03 & 0.32 \\ \hline
Impression Volume & 5.52 & 1.72 & 4.25 & 5.89 & 5.29 & 3.19 & 2.73 & 6.89 \\ \hline
\end{tabular}
\caption{Basic information for all regions and the selected regions. The impression volume's unit is 1000 impressions per day. Bids, reserve prices and budgets are also per 1000 impressions. Active Duration is in days. Bid changes is the average number of agents that change their bid per day in a region. The average of bids, budgets and active duration has been calculated for each agent first and then their averages has been taken over all agents of each region.}
\label{Tab:AllRegionsBasicInfo}
\end{center}
\end{table}
\end{center}
}
In Table \ref{Tab:AllRegionsBasicInfo} we display basic statistics from our data.
The table contrast the statistics for our selected 6 markets with the statistics
of the entire set of 57 markets that we analyzed. Presented statistics
correspond to the number of participating bidders, their bids and budgets,
period of time when the bidder is active in an auction (i.e. has the bid above the
reserve price and did not exhaust the budget), daily frequency of
bid changes and the market reserve prices. The
Table indicates that our selected 6 markets have
similar values of monetary variables (e.g. average bid of 23.9 in selected markets
vs. 18.8 in the entire set of markets and average daily budget of 8.9 in selected
markets vs 9.2 in the entire set of markets).
However, there are two key statistics that are clearly different in our selected set
of markets: the time-average number of participating bidders  (19.3 in selected
markets vs. 10.7 in the entire set of markets) and the average frequency of
bid changes (.7 per day in the selected markets vs. .3 per day in the entire set of
markets).

This means that while the per impression values in our selected markets should be
similar with those in the entire set of experimental markets, our selected markets
have more intense competition and, therefore, we would expect smaller markups
of the bidders and faster convergence of bidder learning towards the optimal bids.

Our data also contains the predicted monthly impression volume for the month ahead (from the start date of each bid).
This estimated impression volume is an input in Zillow's bid recommendation tool whose
goal is to compute the bid that will guarantee that the bidders wins impressions uniformly over time, and wins the maximum expected number of impressions
for the future month for the given budget.
To address the issue of uniform service Zillow implemented budget smoothing explained in the previous section.

An important takeaway from Table \ref{Tab:AllRegionsBasicInfo} is the magnitude of the
relative scale of bids and budgets of bidders across the markets. As it is typical
in display advertising the impression bids are expressed per mille (1000 impressions).
To make the monthly budgets comparable, we convert the budgets to the same scale.
The striking fact is the small scale of budgets relative to the bids. We note that
we computed daily budgets for bidders using the period they were active, which is often only
a subset of time. Note also that for most
bidders these are their true monthly budgets (i.e., they did not increase their budgets to gain more
impressions). This in contrast with the evidence from sponsored search advertising (on Google or Bing) where budgets declared to the advertising platform are often not binding. 
This means that the issue of smooth supply of impressions to each agents
becomes one of the central issues of the platform design. The platform needs to engage
in active management of eligibility of bidders for auction impressions to ensure
that each bidder participates in the auctions at uniform rate over time.

\begin{figure}[!ht]
\begin{center}
  \includegraphics[width=5.0in]{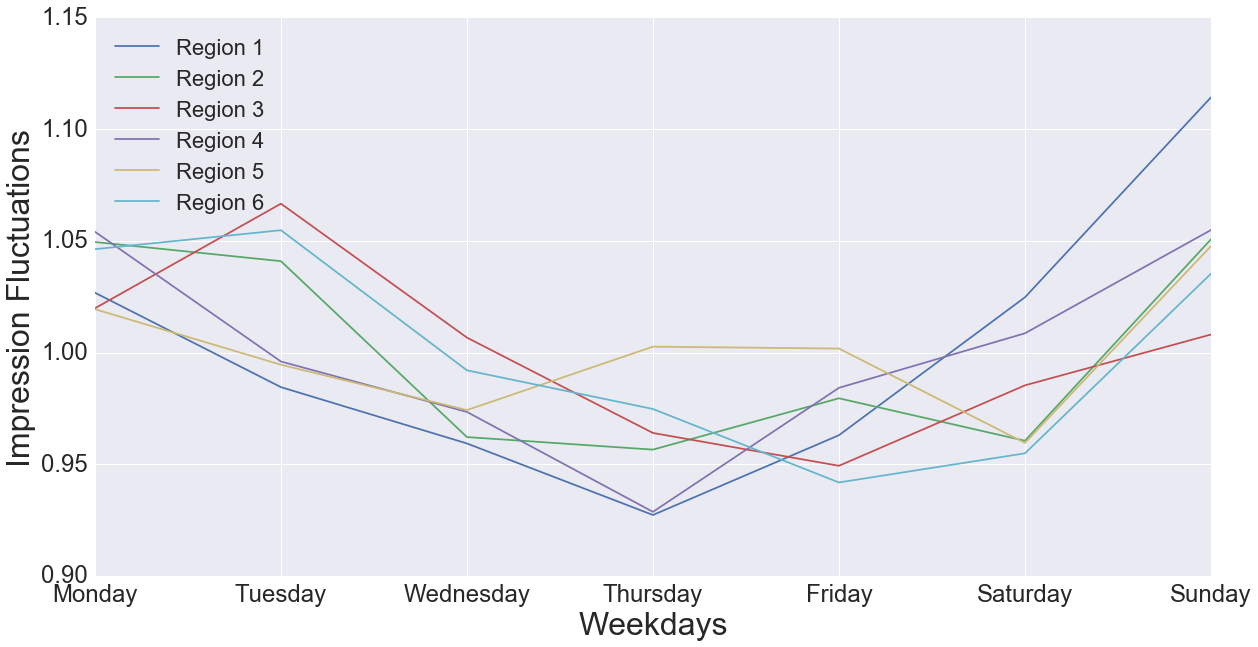}
  \caption{Impression volume fluctuations in weekdays shown for 6 regions with the most number bid changes. The impression volume of each region is normalized by the average daily impression volume of that region.}
  \label{Fig:ImpressionFluctuations}
  \end{center}
\end{figure}

Due to limitations of the data collection, we do not have the data for eligible user impressions
for the entire duration of our auction dataset. In order to properly analyze the auctions, we need the data on the impressions for each bidder for which that bidder was {\it eligible}
including those that she wasn't 
served.
For most of the period we only have Zillow's estimate for user impressions, and only have the actual impression volume for three months. 
For this period 
we noticed that impression volume fluctuates with the days of the week, as shows on Figure \ref{Fig:ImpressionFluctuations}, while the estimated impression volume doesn't show such fluctuation.

To address this data deficiency, we take the predicted volume of eligible impressions
(which is the most reliable proxi for the total number of eligible ad impressions). 
Using the seasonal modulation (mostly reflecting the intra-week changes
of the impression volume) that we observe in the detailed impression
records, 
we augment the impression volume predictions 
to produce a more reliable proxi for daily user impressions. This generates
a realistic pattern for daily impressions for the entire time period that we observe.

\paragraph{\textbf{Data Processing.}}
Zillow's experiments were designed not just to evaluate the performance of auctions
in selected markets per se, but also to compare key characteristics of monetization
and impression \etedit{sales} 
in the incumbent mechanism with negotiated contracts
and the new auction mechanism.
To provide data for credible comparison of these
variables (which we do not analyze in this paper) Zillow did not convert entire markets
to auctions. Instead, a fixed proportion of impressions was reserved for fixed price
contracts and the remaining inventory was released to the auction-based platform.
In each market Zillow selected several agents that were brought to the auction platform
(and who were not allowed to buy impressions from fixed contracts in the same markets).
Towards the end of our period of observation, more agents were getting enrolled in
the auction markets. For most of those new agents the period of observation is too
short for statistically valid inference. As a result, we chose to drop such short-living agents.

\begin{figure}[!ht]
\begin{center}
  \includegraphics[width=5in]{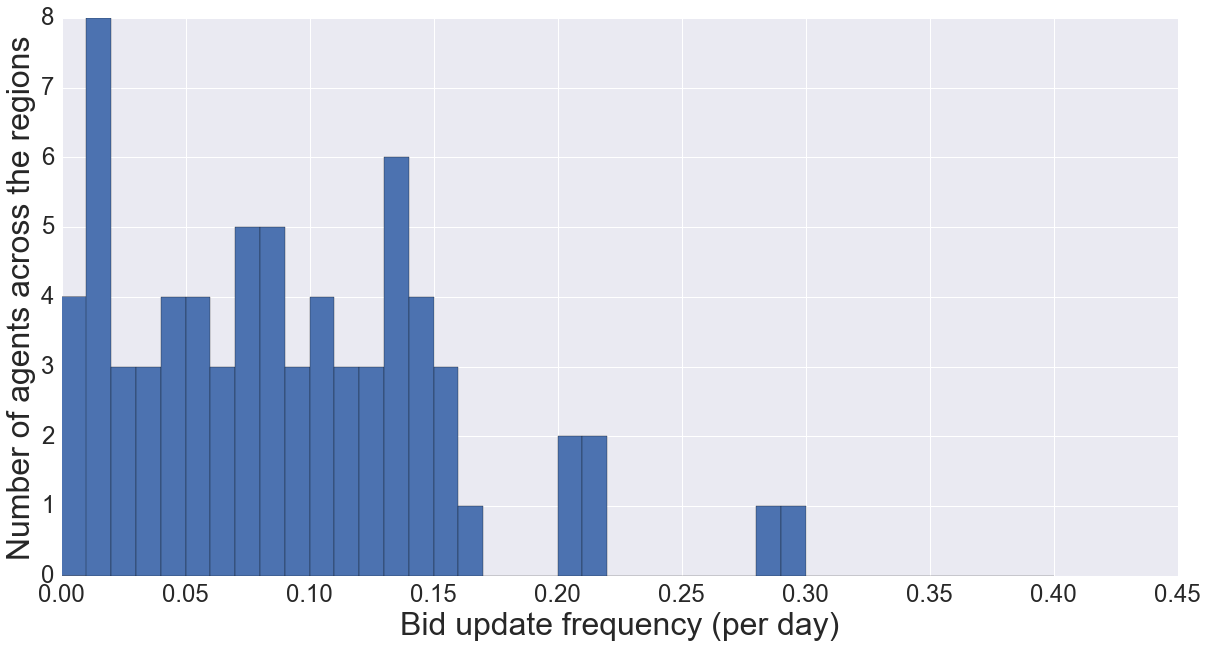}
  \caption{Histogram of bid change frequencies across the 6 selected regions.}
  \label{Fig:HistBidChangeFrequencies}
  \end{center}
\end{figure}

In our structural inference we study the success of agents' bid adjustment over time. We find that a key characteristic of agents is the frequency by which they update bids.
On Figure \ref{Fig:HistBidChangeFrequencies}
we plot the histogram of the distribution of daily frequency of bid changes for all agents across
6 markets. 
The histogram shows fairly spread distribution of frequencies, close
to uniform between the once-a-month update to once a week update with some agents updating the
bids more frequently.

However, the analysis of frequency of bid changes within individual markets
shows much less concordance in the bid update frequency across bidders. In fact the frequency of bid
changes turns out to be the key variable that allows us to cluster the bidders into distinct types. We run k-mean clustering algorithm using the bid change frequency variable to partition the agents inside each region into 3 clusters, agents in cluster 1 and cluster 3 have the lowest and the highest number of bid changes per day respectively. This allows us to give a straightforward interpretation to the types identified by the cluster.
If the bid changes are triggered by the benefit of the bid change out-weighting the
cost of the bid change, then the three clusters can be interpreted as identifying the the bidder with high,
medium and low cost of bid changes.

The results of clustering are demonstrated in Table \ref{Tab:RegionBasicInfo}. The results show fairly balanced
cluster sizes across markets with the high and medium cost clusters containing the largest number of bidders
and the lowest cost cluster containing the smallest number of bidders (less than a quarter of bidders).

{\small
\begin{table}[!ht]
\begin{center}
\begin{tabular}{|l|l|l|l|l|l|l|}
\hline
Region \# &  1 &  2 &  3 &  4 &  5 & 6 \\ \hline
Number of agents & 20 & 23 & 18 & 21 & 16 & 18 \\ \hline
Average bid changes per day & 1.23 & 0.91 & 0.65 & 0.54 & 0.54 & 0.51 \\ \hline
Selected agents in clusters 1,2,3 & 5,5,4 & 5,6,4 & 6,5,1 & 3,2,2 & 8,3,3 & 3,6,1 \\ \hline
\end{tabular}
\caption{Basic information for the 6 selected regions. Filtering removes the agents that are in auction for less than 7 days or do not change their bid at all.}
\label{Tab:RegionBasicInfo}
\end{center}
\end{table}
}

\paragraph{\textbf{Auction simulator.}}
Unfortunately, the system only logged actual delivered impressions for each bidder, and didn't log if a bidder lost the impression due to being outbid, or being filtered, etc. 
Given this limited data, we cannot evaluate system
directly from the data. Instead, we need to simulate it by emulating Zillow's budget smoothing
process. This simulation is the key component of our data processing strategy that will
further allow us to perform structural estimation.

We calculate the outcome of the ad-auctions separately for each region. For each day, we find the set of agents who have active bids in that day, as well as their bid, their daily budget (which is calculated from the monthly budget and the leftover from the previous days), and the region's reserve price. We use this information to simulate the auction for that day by calculating each agent's filtering probability ($\pi_i$), expected payment and expected share of that day's impression volume. 
It is important to note that while we don't analyze data from short lived agents, they are included in the simulation.

One of the main challenges of the system as well as our simulation is to find the filtering probabilities. We describe the details of the algorithm in Appendix \ref{smoothing:appendix}. Recall that the filtering probabilities need to satisfy equation \ref{eq:smoothing}, where $\mbox{eCPM}_i(b_i)$ is a function of the filtering probabilities of other agents. We find an approximate solution to these fixed point equations by minimizing the sum of squares using Newton's method.
$$
\sum^I_{i=1}\left(\pi_i-\min\left\{1,\;\frac{B_i}{\mbox{eCPM}_i(b_i;\pi_1,\ldots,\pi_I)}\right\} \right)^2
$$
with respect to $\pi_1,\ldots, \pi_I$.

The main time consuming step of each iteration is the need to compute the expected cost (eCPM) and expected impressions share (eQ) for all agents with the given probabilities $\pi$. In Appendix \ref{smoothing:appendix} we show how to do this in $O(|I|)$ time (assuming that the bids of the agents are sorted).

\section{Empirical analysis of market dynamics}
\paragraph{\textbf{The dynamics of the adoption of the bid recommendation tool.}}
We study the adoption of
the bid recommendation tool
designed to help the bidders to transition from the
fixed price contracts to the auction-based system for
impression pricing and delivery. The recommendation tool
provided a simple interface that allowed the bidders
to submit their monthly budget for a specific market and
the tool would provide the bid that maximizes the number
of impressions that could be purchased within the month
with the given budget. The tool would adjust the recommendation
with any change that occurs in the system, such as the arrival
of the new bidder, changes of bids by existing bidders, or the change
in the predicted number of market impressions.

During the market rollout Zillow made the agents aware of the
tool's existence and explained the principles that were used
to design the tool. However, despite the outreach and marketing
work when auction platform was introduced to the set of experimental
markets, the actual utilization rate of the tool was initially low.

\begin{figure}[!tbp]
\begin{minipage}[b]{0.5\textwidth}
\begin{center}
  \includegraphics[width=3.2in]{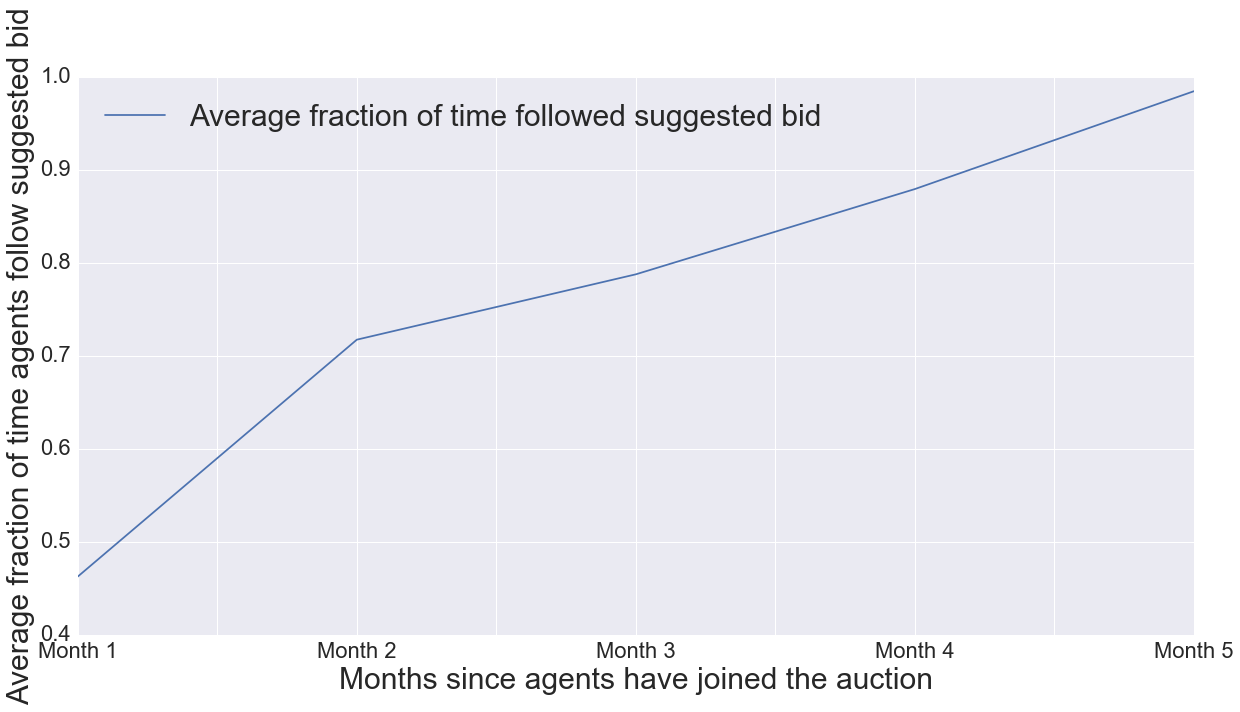}
  \caption{Average fraction of time agents follow the recommended bid across the selected regions.}
  \label{Fig:FollowingSugBidOverTimeAllReg}
  \end{center}
\end{minipage}
\hspace{.5cm}
\begin{minipage}[b]{0.5\textwidth}
\begin{center}
  \includegraphics[width=3.2in]{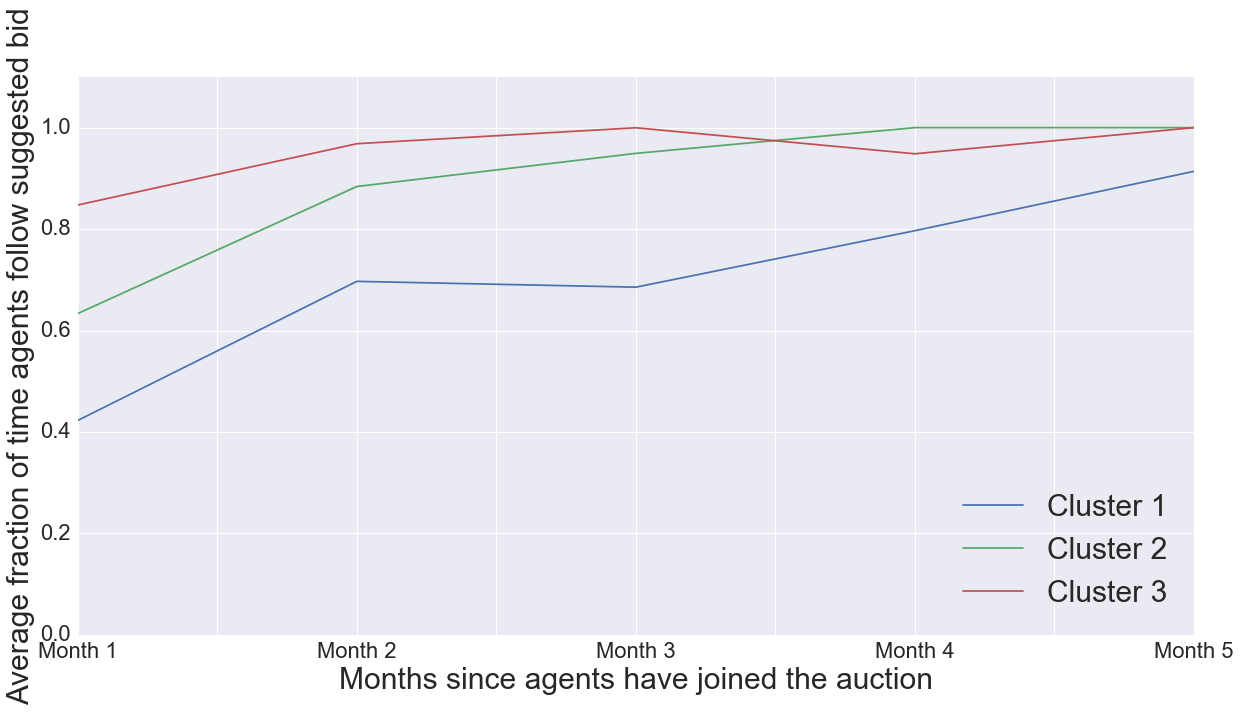}
  \caption{Average fraction of time agents follow the recommended bid separated by clusters.}
  \label{Fig:FollowingSugBidOverTimeClusters}
  \end{center}
\end{minipage}
\end{figure}

On Figure \ref{Fig:FollowingSugBidOverTimeAllReg} we display the percentage of time
when recommended bid was used by a given agent for the bid change over
agent's tenure in the auction platform averaged over all agents in the experimental
markets. The figure shows that when agents were introduced to the platform
they tend to use the recommendation tool for less than 50\% of their bid changes.
This number tends to grow to almost 100\% as the agent is exposed to the
auction platform for more than 5 months.

Figure \ref{Fig:FollowingSugBidOverTimeClusters} presents the same trend
of utilization of the bid recommendation tool but decomposed by clusters.
Recall that 
we cluster bidders based on their bidding
frequency with cluster 1 being the cluster with the lowest frequency of bid changes
and cluster 3 being the cluster with the highest frequency of bid changes.
Figure \ref{Fig:FollowingSugBidOverTimeClusters} shows the persistence of the
trend with the relatively low utilization of the bidding tool when the agents
are just introduced to the auction platform with an increased degree of
this utilization as the agent spends more time bidding in the auctions.
We note that this trend is most rapid for the cluster of bidders with the highest
frequency of bid changes. These bidders start using the bid recommendation
tool for almost all of their bid changes after 3 months of exposure to the
auction platform. At the same time the bidders that are least frequently changing their
bids do not get to the point of fully using the recommendation tool even
after 5 months of experience with the auction platform.

We further illustrate the persistence of this trend across the 6 markets that
we study in the Appendix.
Figure \ref{Fig:FollowingSugBidRegions} in the Appendix confirms the consistency of the aggregate
trend of the utilization of bid recommendation tool with those trends
in individual markets. Moreover,
for some markets the percentage of utilization of the bid recommendation
tool is even smaller than that on average especially for the bidders that change
their bids the least frequently.

This leads us to two important observations. First, the bidders in all observed
markets were willing to ``experiment'' with their bids by deviating from the recommended
bid. The proportion of bids devoted to experimentation is large especially
when the agents are
newly introduced to auction markets and remains large among the bidders
that do not frequently change their bids even after 5 months of them bidding
on the auction platform. Second, even though the bid recommendation tool was designed to
optimize bids on behalf of the bidders, the bidders did not have the full faith that
the recommendations benefit them (as opposed to the auction platform). The increasing
adherence to the recommended bid is observed only after the bidders experiment with
alternative bids for a sufficiently long time.

Next we try to understand if the bidder's learning and experimentation behavior results in improved outcomes, or rather simple helps them trust the platform's recommendation.

\paragraph{\textbf{Trust in system-provided bid recommendations.}}
\label{sec:trust}
To evaluate the bid adjustment in the Zillow's auction markets we use the
methodology developed in \cite{NST:2015}. For understanding the first months of experimenting with auctions, we believe that it is best to model them as off-equilibrium. A characteristic feature of this market that
we analyze in this paper is the relatively small stakes (measured in terms of per impression prices
relative to the budgets). In such markets exploration is a good way to learn the best response.
For agents who change their bid relatively frequently, we model their behavior as {\it no-regret learning} which then allows us to infer their value using the notion of the {\it rationalizable set} from \cite{NST:2015}.
In dynamically changing market bids will vary over time and would not necessarily
maximize utility at each instant. Each bidder will then be characterized by two
parameters: her value and the average regret that evaluates the success of the
dynamic bid adjustment. We measure regret as the 
difference between the time-averaged
utility attained by bidder's bid sequence and the average utility attained by the
best fixed bid in hindsight. Average regret of a player reflects the properties of
bidder's learning algorithm used.

We now consider a dynamic environment where the active bid of the bidder
participates in many auctions for impressions. We assume that time is discrete.
At each instance $t$ bidder $i$ with bid $b_{it}$ and outstanding bids
of other bidders $\vec{b}_{-i,t}$ faces an allocation $eQ(b_{it},\vec{b}_{-i,t};\theta^t)$
and payment  $\mbox{eCPM}(b_{it},\vec{b}_{-i,t};\theta^t)$ produced
by auction outcomes for user impressions that arrived at time $t$.
Here $\theta^t$ are ``environment" variables that reflect time-varying characteristics
such as the rate of arrival of user impressions, budgets and budget smoothing
probabilities.
In our further discussion (where it does not affect mathematical clarity)
we use simpler notation $eQ_{it}(b_{it})=eQ(b_{it},\vec{b}_{-i,t};\theta^t)$
and $\mbox{eCPM}_{it}(b_{it})=\mbox{eCPM}(b_{it},\vec{b}_{-i,t};\theta^t)$
leaving the dependence of allocations and spent from competing bids
and environment variables implicit. Then we can express the utility of
bidder $i$ at instance $t$ as
$$
u_{it}(b_{it},v_i)=v_ieQ_{it}(b_{it})-\mbox{eCPM}_{it}(b_{it}).
$$
The notion of utility allows us to define the average regret of bidder $i$.
\begin{definition}[Average Regret]
A sequence of play that we observe has $\epsilon_i$- average regret for bidder $i$ if:
\begin{equation}\label{eqn:eps-regret}
\forall b'\in {\mathcal B}: \frac{1}{T} \sum_{t=1}^{T}u_{it}(b_{it},v_i) \geq \frac{1}{T} \sum_{t=1}^{T} u_{it}(b',v_i)-\epsilon_i
\end{equation}
\end{definition}

The introduced notion of the average regret leads to the following definition of a \emph{rationalizable set under no-regret learning} (or more precisely, small average regret learning).

\begin{definition}[Rationalizable Set]
A pair $(\epsilon_i,v_i)$ of a value $v_i$ and error $\epsilon_i$ is a rationalizable pair for player $i$ if it satisfies Equation \eqref{eqn:eps-regret}. We refer to the set of such pairs as the \emph{rationalizable set} and denote it with $\NR$.
\end{definition}

\begin{figure}[!th]
\begin{center}
\includegraphics[width=6.5in]{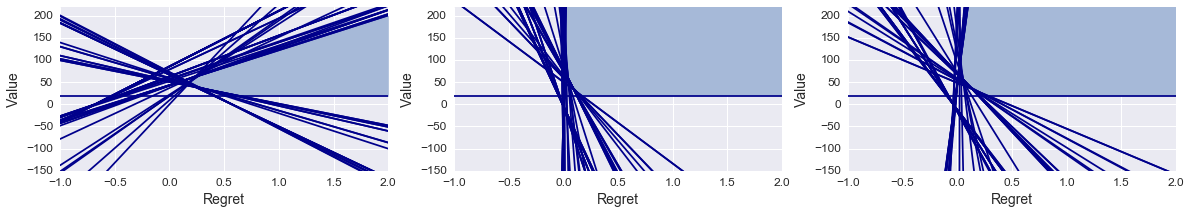}
  \caption{Rationalizable set for 9 agents most frequently changing bids in region 1}
  \label{Fig:HistBidChangeFrequencies:1}
  \end{center}
\end{figure}

To implement the construction of the rationalizable sets we choose a grid over the
bid space and construct half-spaces generated by inequalities (\ref{eqn:halfplanes})
for each bid on the selected grid. On Figures \ref{Fig:HistBidChangeFrequencies:1} we show
the structure of the rationalizable sets for 3 of the bidders most frequently changing their bids
in region 1. The structure of the rationalizable set is similar in all the 6 markets we analyzed, see the corresponding figures \ref{Fig:HistBidChangeFrequencies:12}-\ref{Fig:HistBidChangeFrequencies:56} in the appendix.
The vertical axis on these plots is the per impression value of the bidder (expressed
in monetary units) while the horizontal
axis is the additive average regret.

We note a dramatic difference in the shape of these sets with the rationalizable
sets for the bidders in advertising auctions on Bing estimated in \cite{NST:2015}.
While the rationalizable sets in \cite{NST:2015} have smooth convex shape, the rationalizable
sets for the agents on Zillow are polyhedra.
This is due to a much higher degree
of uncertainty for the bidders on Bing (induced by variation of estimated clickthrough rates
and user targeting across user queries) that smooths out the boundary of the
rationalizable set.

Another important observation is the highly concentrated set of hyperplanes that
pass through the origin for many the bidders in the observed markets.
In fact, as we mentioned before, agents' budget constraints on Zillow are
binding for most bidders with the per impression bids exceeding the
per impression budgets. This means that for those budget constrained bidders
there is a set of bids that correspond to them completely spending the budgets
(i.e. that all have identical expected spent). Thus regret of these agents
is determined only by how many impressions each fixed bid with a binding budget generates but not their spent implying that
$
v_i \frac{1}{T}\sum^T_{t=1}\left(eQ_{it}(b)-eQ_{it}(b_{it})\right)\leq \epsilon_i,
$
which for each bid $b$ is the half-space that contains the origin.
For bidders who spend their budgets, the small regret constraint of the rationalizable set does not give any upper bound on their valuation: these bidders are constrained by their budget, and not by their value of each impression.

In our further analysis we focus on the specific point of the rationalizable
set that corresponds to the pair of value and regret where the observed bid sequence
has the smallest possible average regret. Since the average additive regret and the value
are expressed in the same monetary units, we can directly compare them.
A simple visual analysis of plots of rationalizable sets on Figures \ref{Fig:HistBidChangeFrequencies:12}-
\ref{Fig:HistBidChangeFrequencies:56}  indicates that while for some bidders the smallest rationalizable
average regret is small relative to the corresponding value, there is a large number of bidders
with high relative regret. This is particularly pronounced for the bidders with cone-shaped
rationalizable sets. From the economic perspective, this shape indicates that a small change
in the bid for those bidders from the applied bid would have resulted either in a significant
increase in the number of allocated impressions or in a significant drop in the per impression cost.

We want to understand why agents may not be following the platform provided recommendation: do they use a different bidding strategy as that improves their obtained utility, or is it simply a question of lack of trust? We note that the bid-recommendation tool didn't take into account the weekly impression volume fluctuation shown on the figure in Section 3, so with bidding differently on weekdays and weekends, the agents could have done better than the recommendation, and in fact would be able to achieve negative regret.

To evaluate regret, we need to infer the agents value for the impression. The rationalizable set offers a convex set of possible value and regret pairs.
We use the value with smallest rationalizable additive regret as our selected value,
but to display the values in context, we also want to account for two features.
First, it useful to measure regret relative
to bidder's value (i.e. the bidders may be prone to evaluate the ``loss" associated with their
learning strategies in the increments of the total ``gain"). Second, it is also
convenient to normalize the regret by the the number of impressions the agent won,
so we measure ``per impression'' regret.


In Table \ref{Table:RegretDiff2} we show summary statistics on whether agents would
decrease or increase their regret by not using a platform recommended recommendation in each of the 6 markets. All regrets are computed using the value we inferred using the agent's own bid.
Whenever we say that the regret of a bidder's learning strategy is the same as the regret of the recommended bid
either her own learning
strategy is as good as the recommended bid or that she simply adhered
to the recommended bid. The overall distribution of the regret difference indicates that sizeable fractions of bidders both have the regret
that exceeds the regret of the recommended bid and the regret that
is smaller than that if the recommended bid was used. We also show the same statistics
by regions and clusters, where cluster 3 are the agents who update their bids most frequently.

{\small
\begin{table}
\begin{center}
\begin{tabular}{|l|l|l|l|l|l|l||l|l|l||l|}
\hline
Region & reg 1 & reg 2& reg 3 & reg 4 & reg 5 & reg 6 & cl 1 & cl 2 & cl 3 & all\\
\hline
 worse & 21.4\% & 20\%& 8.3\% &0\% & 21.4\% & 50\% &23.3\% & 25.9\%&  6.7\%& 20.8\%\\
better& 42.9\% & 20\%& 41.7\% & 28.6\% & 21.4\% & 20\%& 46.7\%& 22.2\%&6.7\% & 29.2\%\\
equal & 35.7 \%& 60\%& 50\% & 71\%& 57.1\%& 30\% & 30\% & 51.8\% & 86.7\%& 50\%\\
\hline
\end{tabular}
\end{center}
\caption{The percentage of agents that do worse (or better) with their bids than following the recommendation. The three columns on the right of the table offer aggregate statistics across the 6 markets segmenting agents by the frequency they update.}
\label{Table:RegretDiff2}
\end{table}
}

\begin{figure}[!th]
\center
\includegraphics[width=6.5in]{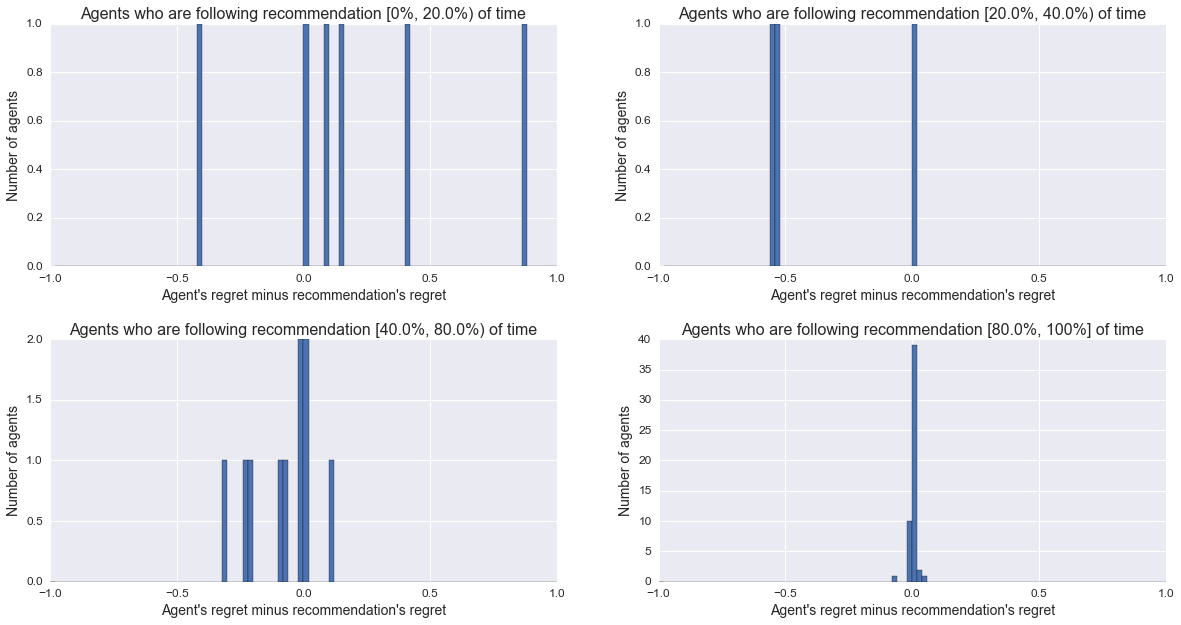}
\caption{Distribution of difference between the regret of own bidding
strategy and recommended bid across agents in selected markets separated by the percentage of time the agent follows the recommendation.}
\label{Regret:Distribution}
\end{figure}

We further illustrate this point on Figure \ref{Regret:Distribution}.
The distribution of differences between the regret of agents' own bidding
strategy and the recommended bid is close to symmetric about zero
for all markets that we study. This indicates that even though the bidders
chose to experiment with their bids, the experimentation did not
necessarily lead to an improvement of regret over the recommendation.
In fact, about a half of the bidders have worse regret from their deviating
strategy than they would have had if they always chose the recommended bid.

\begin{figure}[!th]
\center
\includegraphics[width=5in]{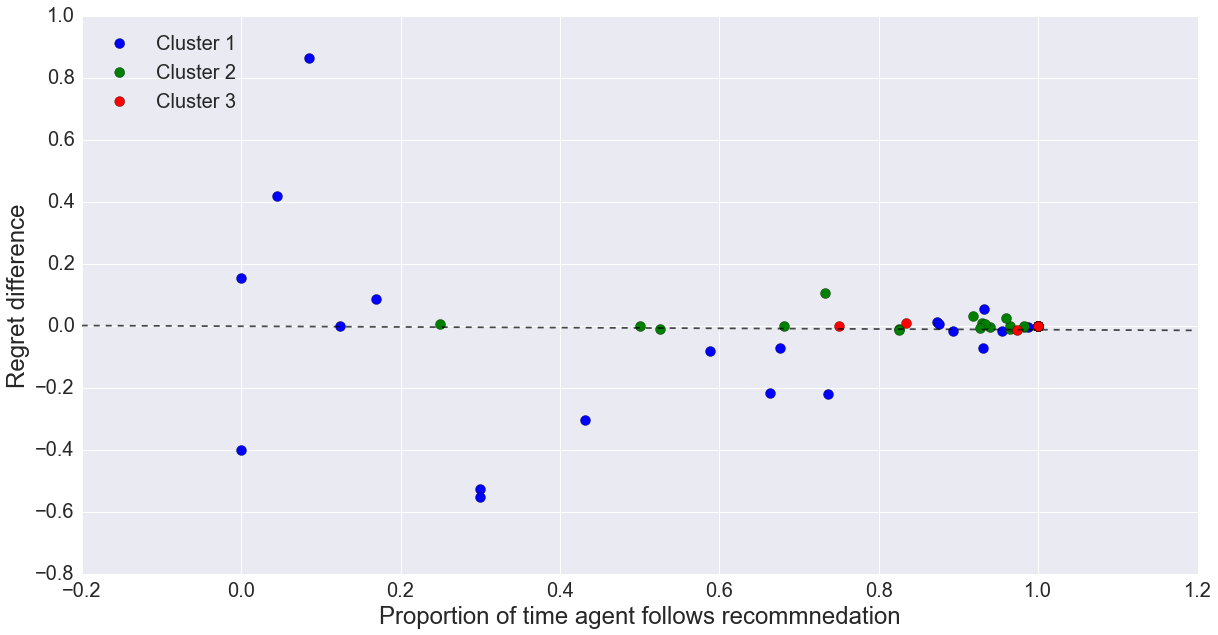}
\caption{Scatter plot of difference between the regret of own bidding
strategy and recommended bid across agents and the percentage
of time agents use the recommended bid. The dashed
line shows the best linear fit.}
\label{Regret:Scatter}
\end{figure}

One a priori possible explanation for why agents don't follow the platform recommendation
is that the recommended bid does not provide satisfactory outcomes
for the agents and switching to an alternative bidding sequence improves
their long-run performance.
Figure \ref{Regret:Scatter} shows that this hypothesis is not consistent
with the data. On average, the agents who use the recommended bid less do not show
any improvements over the recommended bid measured by the average regret.

Combining this information with the previous observation of an increasing trend
of utilization of bid recommendation
tool, we conclude that the key element that explains our results is the {\it trust}
of the agents in the platform-provided bid recommendations. While the bid
recommendation tool is optimized for the agents, upon entry to the platform
the agents do not trust the tool. Instead, they experiment with alternative
bids and compare the performance of those deviating bids with the performance
of the recommended bids that they also occasionally choose. Once the agents
empirically verify that the tool indeed optimizes the bids on their behalf, they
start using the tool for most of the bid changes.

\paragraph{\textbf{Conclusion.}}
Our conclusion is that the agents have sacrificed the performance of
their advertising budgets over a long period of time just to ensure that
choosing the recommended
bid over some other alternative bid does not make them worse off. In this
case, it would have been optimal both from the perspective of the long-term
welfare of the agents and from the perspective of stability of prices
on the platform to simply default all bidders to recommended bids.  


\bibliography{Sections/bibliography_Zillow}

\appendix
\section*{Appendix}

\section{Algorithmic description of computation of budget-smoothing probabilities. }\label{smoothing:appendix}
The following steps outline our budget smoothing algorithm:
\begin{enumerate}
\setlength{\itemsep}{0pt}\setlength{\parsep}{0pt}\setlength{\parskip}{0pt}
\item Sort the bidders $i$ by their bid $b_i$ and assume bidders are numbered in this order.
\item Construct an array of $2^I$ binary $I$-digit numbers from $\{0,0,\ldots,0\}$ to $\{1,1,\ldots,1\}$\etedit{where the number $N=n^N_1n^N_2 \ldots n^N_i\ldots n^N_I$ will correspond to bidders $j$ with $n^N_j=1$ not being filtered out}. Call the set of elements in this array $\mathcal N$
\item Take a subset of elements of $\mathcal N$ where $i$-th digit is equal to 1. Call this set ${\mathcal N}_i$
\item Let $N=n^N_1n^N_2 \ldots n^N_i\ldots n^N_I$ with $n^N_i=1$ and $n^N_j \in \{0,1\}$ be a specific row in ${\mathcal N}_i$, \etedit{corresponding to the outcome of filtering when the agents with $n_j^N=1$ remained}.
\item Include the bid of each bidder $j$ for whom $n^N_j=1$,and determine the price of bidder $i$, calling it $\mbox{PRICE}^N_i$\etedit{which is the maximum of the reserve price, and the bid $b_j$ first agent $j>i$ with $n^N_j=1$. Let $j^N_i$ be the position of agent $i$ after filtering, and let $\gamma_j^N$ the corresponding probability $\gamma_{j^N_i}$}.
\item Compute the expected spent as
$$
\mbox{eCPM}_i(b_i;\pi_1,\ldots,\pi_I)=\sum\limits_{N \in {\mathcal N}_i}\etedit{\gamma_i^N} \prod_{j \neq i} \pi_j^{n^N_j}(1-\pi_j)^{\etedit{(1-n^N_j)}}\,\mbox{PRICE}^N_i.
$$
\item Solve for $\pi_1,\ldots, \pi_I$ by solving a system of nonlinear equations
$$
\pi_i=\min\left\{1,\;\frac{B_i}{\mbox{ eCPM}_i(b_i;\pi_1,\ldots,\pi_I)}\right\},\;\;i=1,\ldots,I.
$$
For instance, we can find an approximate solution by minimizing the sum of squares \etedit{using gradient decent or Newton's method}.
$$
\sum^I_{i=1}\left(\pi_i-\min\left\{1,\;\frac{B_i}{\mbox{eCPM}_i(b_i;\pi_1,\ldots,\pi_I)}\right\} \right)^2
$$
with respect to $\pi_1,\ldots, \pi_I$.
\end{enumerate}

The main part of the above iterative algorithm for finding a fixed point is to compute the expected cost (eCPM) and expected impressions share (eQ) for all agents for a given set of probabilities $\pi_1,\ldots,\pi_I$. Considering all subsets of agents, this
can take exponential time in number of agents. We to to run the simulations for every day and each region separately and we need to compute these many times to get the filtering probabilities, these calculations can significantly increase the running time of our simulations. Furthermore, for auctions where the number of agents is big, using an exponential time algorithm to calculate the outcome of each iteration is not feasible. In order to get around this issue, we use the fact that in each underlying GSP, only impressions of the first four agents are eligible to be shown, as well as the fact that the filtering probabilities of agents are independent. Our algorithm to find expected cost (eCPM) and expected impressions share (eQ) for given filtering probabilities runs in linear time (in number of agents).

For each agent $i\in[I]$, our algorithm first computes the expected impression share of the agent $eQ_i$ (assuming she has not been filtered). Note that $eQ_i$ only depends on the number of unfiltered agents ($r$) that have a higher bid than $i$ and the \etedit{probability ($\gamma_{r+1}$) associated with} $i$'s rank.
We first find the probability that there are exactly $r\in [0,3]$ number of agents with higher bid than $i$, called $p_{i,r}(\pi)$. By multiplying $p_{i,r}(\pi)$ by the impressions of the $(r+1)$-th position ($\gamma_{r+1}$) we can find the expected impression share of agent $i$.

After finding the expected impression share of $i$, we calculate $eCPM_i$ by using $eQ_i$. In order to do this, we use the fact that the filtering probabilities of agents are independent. Furthermore, the expected cost per impression for agent $i$ is only a function of bids of agents who are bidding lower than $i$ and their filtering probabilities. So by calculating the expected cost per impression and multiplying it by the expected impression share of agent $i$, we can calculate her expected payment conditioned on $i$ being in the auction ($eCPM_i$). Recall that for each agent $i$, $eQ_i$ and $eCPM_i$ are calculated conditioned $i$ not getting filtered, so in order to calculate the total expected number impressions that she wins in the auction and her expected spent, it is enough to multiply $eQ_i$ and $eCPM_i$ by $\pi_i$. In algorithm 1 we have marked these steps for each agents.

{\small
\begin{algorithm}
\label{Alg:eCPM}
\DontPrintSemicolon
\KwIn{$b$: bids (sorted) , $\pi$: filtering probabilities, $\gamma$: rewards, $reserve$: the reserve price}
\KwOut{$eCPM,eQ$}
Let $\{1,2,\ldots,I\}$ be the list of all agents such that $b_1\geq b_2\geq \ldots \geq b_I \geq reserve$\;
Let $p_{1,0}(\pi)=1$ and $p_{1,r}=0$ for $0<r\leq 3$\;
\For{$i \in [I]$}{
\If{$i>1$}{
	Let $p_{i,0}(\pi)=(1-\pi_{i-1})p_{i-1,0}(\pi)$\;
\smash{\makebox[12.5cm][r]{$\left.\begin{array}{@{}c@{}}\\{}\\{}\\{}\\{}\end{array}\right\}%
\begin{tabular}{l}Calculating $p_{i,r}(\pi)$\\from $p_{i-1,r}(\pi)$\end{tabular}$}}
	\For{$r \in [1,3]$}{
		Let $p_{i,r}(\pi)=(1-\pi_{i-1})p_{i-1,r}(\pi) + \pi_{i-1}p_{i-1,r-1}(\pi)$
	}
}
Let $eQ_i=0$\;
\For{$r \in [0,3]$}{
\smash{\makebox[11.5cm][r]{$\left.\begin{array}{@{}c@{}}\\{}\\{}\end{array}\right\}%
\begin{tabular}{l}Calculating expected impression share \\ $eQ_i$from $p_{i,r}(\pi)$\end{tabular}$}}
$eQ_i=eQ_i+\gamma_{r+1}.p_{i,r}(\pi)$\;
}
Let $j=i+1$\;
Let $CPM_i=0$\;
\uIf{$i=1$ or $\pi_i=1$}
{
\While{$j\in [I]$ and $\pi_j<1$}{
\uIf{$j=i+1$}{
$q_{i,j}(\pi)=\pi_{j}$\;
}\Else{
$q_{i,j}(\pi)=\frac{\pi_j(1-\pi_{j-1})}{\pi_{j-1}}q_{i,j-1}(\pi)$\;
}
$CPM_i=CPM_i + b_j.q_{i,j}(\pi)$\;
$j=j+1$\;
}
\smash{\makebox[11.5cm][r]{$\left.\begin{array}{@{}c@{}}\\{}\\{}\\{}\\{}\\{}\\{}\\{}\\{}\\{}\\{}\\{}\\{}\\{}\\{}\\{}\\{}\\{}\end{array}\right\}%
\begin{tabular}{l}Calculating cost per impression\end{tabular}$}}
}\Else{
$CPM_i=\frac{CPM_{i-1} - \pi_{i}b_{i}}{1-\pi_{i}}$\;
}
\If{$j=I+1$}{
\uIf{$i=I$}{
$q_{i}(\pi)=1$\;
}\Else{
$q_{i}(\pi)=\frac{1-\pi_{j-1}}{\pi_{j-1}}q_{i,j-1}(\pi)$\;
}
$CPM_i=CPM_i + reserve.q_i(\pi)$\;
}
$eCPM_i= eQ_i.CPM_i$
}
\caption{Calculating eCPM and $eQ$ of agents in linear time.}
\Return $eCPM,eQ$
\end{algorithm}
}

If this algorithm is implemented naively, the most expensive computation is computing $p_{i,r}(\pi)$ for all the agents. For each agent, it takes $O(I^3)$ to find all the configurations where there are $r\in [0,3]$ agents who are not filtered and have higher bid than $i$. For each configuration, it takes $O(I)$ to compute its probability. By using dynamic programming, we can compute $p_{i,r}(\pi)$ from $p_{i-1,r}(\pi)$ by considering the cases where agent $i$ is getting filtered and is not getting filtered separately. For initialization, we set $p_{1,0}(\pi)=1$ and $p_{1,r}(\pi)=0$ for $0<r\leq 3$. We use the following update rule to computer $p_{i,r}(\pi)$ for $i>0$:
$$
p_{i,r}(\pi)=\begin{cases}
(1-\pi_{i-1})p_{i-1,r}(\pi) & r=0\\
(1-\pi_{i-1})p_{i-1,r}(\pi) + \pi_{i-1}p_{i-1,r-1}(\pi) & 0<r \leq 3
\end{cases}
$$

This reduces the running time of computing $p_{i,r}$ for each agent from $O(I^4)$ to $O(1)$. The calculations for finding $CPM$ and $eCPM$ can be done in a similar way: instead of computing eCPM for each agent from scratch, we can compute the expected price per impression from the previous calculations. When $i=1$ (she is the highest bidder), or $\pi_i=1$ (she is never getting filtered), first, we set the expected cost per impression to 0 and find the smallest $j>i$ such that $\pi_j=0$. Then, for each $k\in (i,j]$, we find the probability that all the agents $z \in (i,j)$ are getting filtered and $k$ is not getting filtered ($q_{i,k}(\pi)$), multiply it by bid of $k$ ($b_k$) and add the result number to the expected cost per impression. If for all $j>i$, $\pi_j<1$ then we set $j=I+1$ and we assume that $I+1$ is an agent who is never getting filtered and has a bid equal to the reserve price. Note that we also compute $q_{i,k}(\pi)$ from $q_{i,k-1}(\pi)$ in $O(1)$, instead of computing it for each $k$ from scratch.

When $i>1$ and $\pi_i<1$, we use the previous expected cost per impression that we had from the previous agent (lets call it $CPM_{i-1}$) to calculate the cost per impression of agent $i$ ($CPM_{i}$) by doing the following calculation
$$CPM_{i}= \frac{CPM_{i-1} - \pi_{i}b_{i}}{1-\pi_{i}}$$
This operation nullifies the effect of agent $i$ in the cost per impression of the previous agent ($i-1$) and calculates the new expected cost per impression. Finally, we set $eCPM_i=eQ_i.CPM_i$. Note that even though the running time of this algorithm may be $O(I)$ for some agents, the total (amortized) running time of these calculations for all the agents combined is still $O(I)$. So overall the algorithm requires amortized $O(1)$ number of calculations for each agent and it takes linear time ($O(I)$) to calculate eCPM and eQ for all the agents, given the sorted list of agents based on their bids. Since we need to sort the agents by their bids at the beginning, the total running time of each iteration in computing the filtering probabilities is $O(I \log(I))$. This improvement in the running time (from $O(I2^I)$ to $O(I)$) is crucial for simulating the outcome of the auction, specially in the auctions where the number of agents is big.

\section{Bid recommendation tool.}
To help the advertisers, the platform also provides a bid-recommendation platform-provided
bid recommendation. The agents in this market are real-estate agents, who often don't have the data or the analytic tools to do a good job optimizing their bid. 
In addition, the market participants are real estate agents
for whom Zillow may not be the main channel through which they get the ``client leads''.
As a result, some agents may be reluctant to engage in active exploration of optimal
bidding in the auction market for user impressions. In order to facilitate the
bidding for those agents, the platform has developed a tool that recommends
the bid for a given bidder based on this bidder's monthly budget.
The tool was designed to set the bid that maximizes the expected number of impressions
that a given bidder gets given her budget. We now outline the design details of this tool.

\begin{figure}
\center
 \begin{minipage}{.48\textwidth}
  \includegraphics[width=3in]{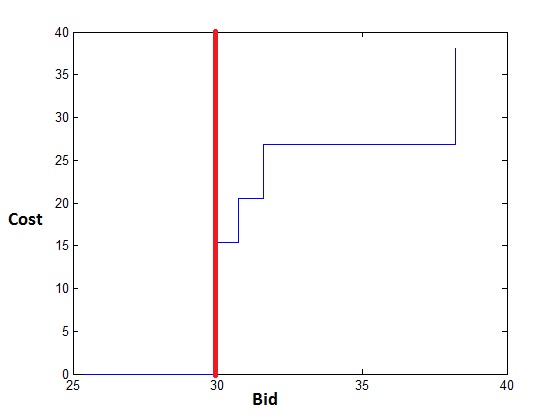}
  \caption{Spent (agent's bid in red)\label{fig2}}
 \end{minipage}
 \begin{minipage}{.48\textwidth}
  \includegraphics[width=3in]{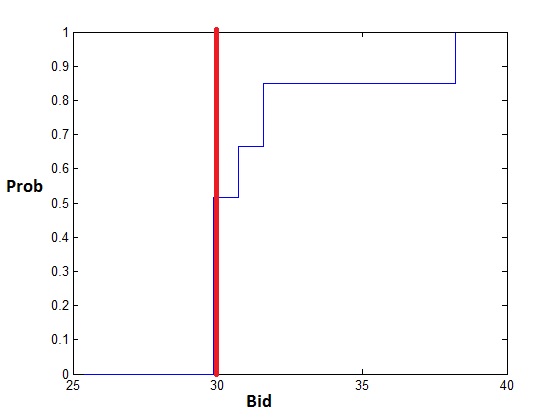}
  \caption{Impression share (agent's bid in red)\label{fig3}}
 \end{minipage}
\end{figure}

For each actual realization of the group of competing bidders, we
define the cost function $CPM_j(b_j)$ as a mapping from the bid
of bidder $j$ to the price she pays per impression in an auction. Recall that we
defined $Q_j(b_j)$ as a probability of being allocated
an impression as an outcome of an auction conditioned on that the agent wasn't filtered.
Note that without the effect of filtering both functions are step functions: whenever bidder $j$ outbids
bidder ranked $i$ but ranks below bidder ranked $i-1$, then this bidder
$j$ pays the price determined by $i$ $b_j$ between $b_i$ and $b_{i-1}$.
In figures \ref{fig2},\ref{fig3} we illustrate the concept of price
and the impression probability using the bidders in one of our 6 markets
and take the bidder ranked 4 in the first week in the market and bid of 30 (recall that the
price units were resealed not to reflect the actual market prices). 
The figure shows
the price and the impression share that this bidder gets if all other
bidders are made eligible for this impression.

We note that the bidder under consideration has a per impression
budget significantly below the cost of an impression (as much as a factor of 10 below).
If the eligibility status for impressions was recorded
by the system, then we could compute the empirical fraction of impressions where this
given bidder was made eligible for an auction: divide the number of impressions where
a given bidder was made eligible for an auction by the total number of
arrived impressions. Depending on the impression volume, this can be computed
using all impressions from the beginning of the month or some smaller window of time
(e.g. the week before). This would be our estimated probability $\pi_j$ of actually getting displayed for an impression.

\paragraph{\textbf{Expected spent and expected impression allocations}}
If many bidders are affected by the budget
smoothing than the participation of bidders in an auction is random (where randomness is
activated by the budget smoothing mechanism). The for each impression instead of the actual spent and the impression share we will have the expected spent and the impression share, where the expectation is
taken with respect to the randomness of participation of competing bidders.
Given the filtering probabilities, the expected spent and expected impression share
are represented by the expectation of spent and impression shares in all possible
bidder configurations which are then weighted by the probabilities
of those configurations. The participation of each bidder $i$ in a given auction
can be represented by a binary variable where $1$ indicates that the bidder
is made eligible for the auction and $0$ means that the bidder was made ineligible due
to budget smoothing. Then the set of all possible participating bids can be represented
by an array of $2^I$ binary $I$-digit numbers from $\{0,0,\ldots,0\}$ to $\{1,1,\ldots,1\}$. Call the set of elements in this array $\mathcal N$. Then a subset of elements of $\mathcal N$ where $i$-th digit is equal to 1
corresponds to the configurations where bidder $i$ is made eligible for an auction.
Call this set ${\mathcal N}_i$. Then the expected cost and the expected impression share are
computed as
\begin{equation}\label{eCPM}
eCPM_i(b_i)=\sum\limits_{N \in {\mathcal N}_i}\gamma_i^N\prod_{j \neq i} \pi_j^{n^N_j}(1-\pi_j)^{n^N_j}\,\mbox{PRICE}^N_i(b_i),
\end{equation}
and
$$
eQ_i(b_i)=\sum\limits_{N \in {\mathcal N}_i}\prod_{j \neq i} \pi_j^{n^N_j}(1-\pi_j)^{n^N_j}\,\gamma_i^N,
$$
where $N$ corresponds to the index of the set of eligible of bidders.

\paragraph{\textbf{Computation of filtering probabilities}}
If the impression allocations are not available, then the probabilities of participation
of bidders in impressions have to be computed. We can consider the actual budget smoothing
as an iterative process: we continuously evaluate the actual spent for each bidder and when
the spent exceeds the allocated budget, then the bidder is made ineligible for some
impressions. This iterative process reaches the steady state when the expected spent in an
impression for a given bidder becomes equal to the budget:
$$
\pi_i \times  eCPM_i(b_i)=\mbox{Budget}_i.
$$
We can simulate this iterative process for the bidder in one of our selected markets.
Note that that bidder has a very low per impression budget of 3.87.
We start the process assuming that all the bidders are eligible for an impression. Then
using the spent in that impression, we compute the filtering probabilities
for all by dividing the budget by the spent and then iterate the process to set
$$
\pi_i=\frac{\mbox{Budget}_i}{eCPM_i(b_i)}
$$
using the previous iteration values of the eligibility probabilities.
The algorithm for computing the probabilities of being displayed on the
page is the following.

\paragraph{Iteration 0:} Initialize probabilities of being eligible for an impression
at $\pi_{i}^{(0)}=1$.
\paragraph{Iteration k:} Take the probabilities of being eligible for an impression
$\pi_{i}^{(k-1)}$ computed from the previous iteration. Compute eCPM
from (\ref{eCPM}) for each bidder $i=1,\ldots,I$. If $eCPM_i(b_i)=0$, then
set the probability $\pi_i=1$ (bidder is always displayed, this bidder never
gets any impressions as an outcome of the auction).
If $eCPM_i(b_i)>0$, then set
$$
\pi_{i}^{(k)}=\min\left\{1,\,\frac{\mbox{Budget}_i}{eCPM_i(b_i)}\right\}.
$$
\paragraph{Stopping criterion:} Stop when the probabilities become close
across the iterations: $\max_i|\pi_{i}^{(k)}-\pi_{i}^{(k-1)}|<\epsilon$, for a
given tolerance criterion.

We illustrate the trajectory across the iterations the bidder of interest
on the following figure.

\begin{figure}[!ht]
\center
 \begin{minipage}{.32\textwidth}
  \includegraphics[width=2in]{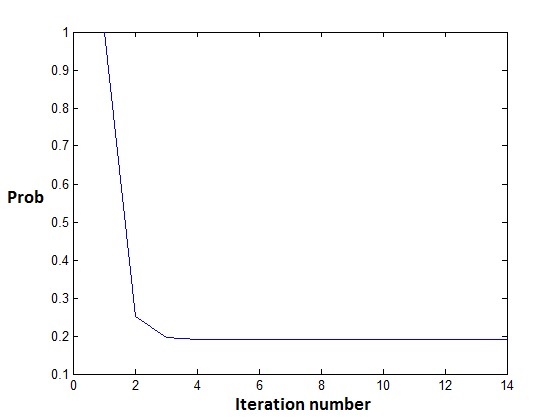}
 \end{minipage}
 \begin{minipage}{.32\textwidth}
  \includegraphics[width=2in]{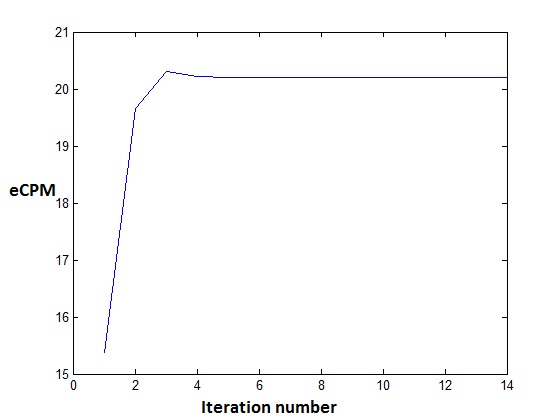}
 \end{minipage}
  \begin{minipage}{.32\textwidth}
  \includegraphics[width=2in]{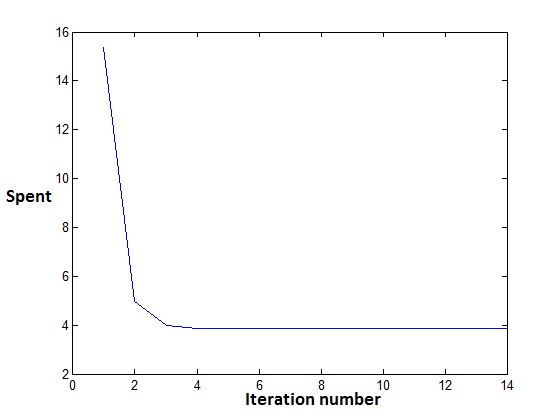}
 \end{minipage}
\caption{Iteration path for probability of eligibility, eCPM and expected spent}
\end{figure}
At the end of the iterative process the expected spent approaches the budget
due to the increase in the filtering probability. Note that the expected CPM
significantly increases in response to the change in the filtering probabilities for
all other bidders.

The randomness increases the eCPM and the probability of allocation into an impression
as compared to the fully deterministic case (i.e. when
bidders are not randomly removed from impressions due to budget smoothing). The figure below
demonstrates the expected CPM and expected fraction of impressions (after budget smoothing)
for bidder with bid of 30 and per impression budget of 3.87.

\begin{figure}[!ht]
\center
 \begin{minipage}{.49\textwidth}
  \includegraphics[width=3in]{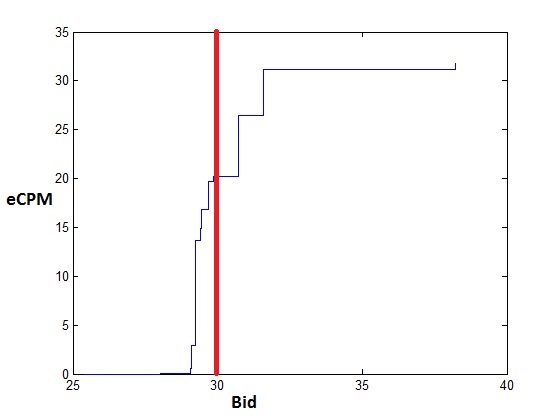}
  \caption{Spent (actual bid in red)}
 \end{minipage}
 \begin{minipage}{.49\textwidth}
  \includegraphics[width=3in]{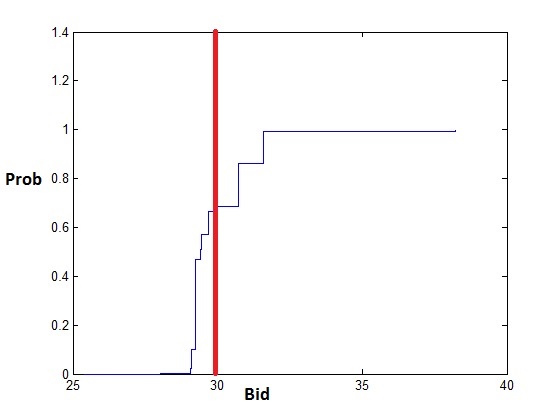}
  \caption{Impression fraction (actual bid in red)}
 \end{minipage}
\end{figure}

Note that an increase in the auction outcomes (probability of being allocated an
impression and the eCPM is compensated with a decrease in the probability of
being eligible for an auction).

\paragraph{\textbf{Computation of the optimal bid for impression ROI optimizers}}
We note that the eCPM and allocation probabilities are monotone functions
of the bid. As a result, if a given bidder maximizes the probability of appearing
in an impression as a function of the bid, the optimal bid will be set such that (a) the expected
spent does not exceed the per impression budget;  (b) an increase in the bid will result
in an increase in the spent exceeding the budget. Note that the spent is (non-strictly)
monotone increasing until it reaches the level of the per impression budget and then it stays
at the level equal to the budget due to budget smoothing. Assuming that the bidders do not
have the ``values of residual budget", this means that the bidder whose budget per impression
exceeds any other budgets should set the bid at the level equal to the budget. Note that the deviation from this strategy will not be optimal: a decrease in the bid for such a bidder results in ``budget savings" that have no value for this bidder, but at the same time it will result in a (weak) decrease in the number of impressions.

The tool that computes the optimal best response for such a bidder proceeds in the following way.

\paragraph{Construction of the grid of bids} We construct the grid of bids of opponent bidders. These are the points where the spent function exhibits jumps.

\paragraph{Construction of the eCPM curve} We construct the eCPM curve. By choosing a small $\epsilon$ (smaller than the minimum distance between the closest score-weighted bids), we evaluate the changes in the eCPM after a given bidder outbids and under-bids the opponent by $\epsilon$.

\paragraph{Computation of the optimal bid} Set the bid to the level where the eCPM curve intersects the horizontal line corresponding to the budget.

\paragraph{Adjustment of the bid for top/bottom bidders} The top bidder sets the bid at the level equal to the per impression budget, the bottom bidder sets the bid to the maximum level that makes the spent positive.

Note that if there are $I$ bidders, this approach amounts to $2\times I$ evaluations of the eCPM function.
The picture below demonstrates the shift to the optimal bid for the bidder under consideration by equating this bidder's eCPM with the per impression budget.

\begin{figure}[!ht]
\center
\begin{minipage}{.49\textwidth}
  \includegraphics[width=3in]{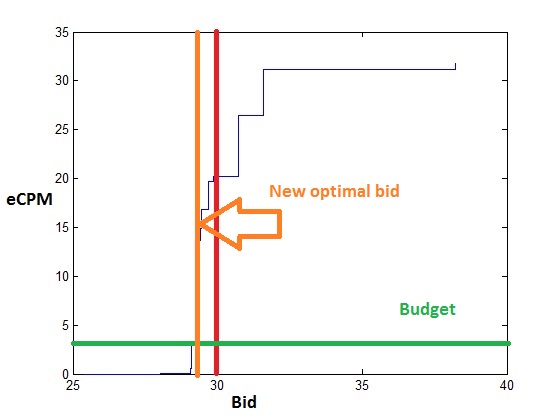}
  \caption{Optimization of impression ROI}
\end{minipage}
\begin{minipage}{.49\textwidth}
  \includegraphics[width=3in]{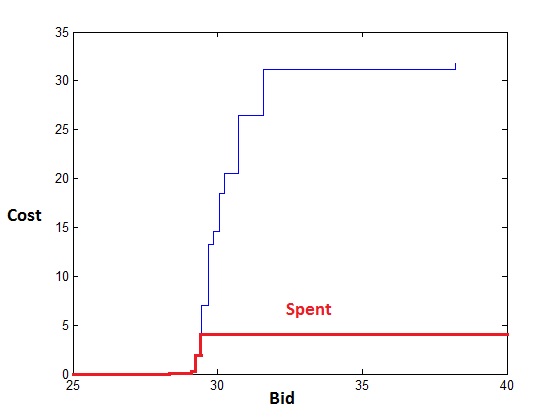}
  \caption{The impact of budget smoothing on expected spent}
\end{minipage}
\end{figure}

We note what happens to the actual spent of the bidder whenever the bid exceeds the recommended level.
Given that at the recommended level the bidder's spent is at or below the per impression budget,
if the bid increases then the budget smoothing gets initiated. The overall spent per
impression is equal to the product of the probability of being eligible for an auction ($\pi_i$)
and the expected outcome of an auction ($eCPM_i(b_i)$)
$$
\pi_i\,eCPM_i(b_i).
$$
Thus, whenever the budget smoothing is initiated ($\pi_i<1$) then the spent is exactly equal to the budget.
Thus the spent as a function of the bid will become flat once the optimal bid has been exceeded.

The probability of being allocated an impression is equal to the product of the probability
of being eligible for an auction ($\pi_i$) and the probability of being allocated an impression
as an outcome of an auction ($eQ_i(b_i)$). We note that since the GSP is monotone, the
probability of being allocated an impression as an outcome of an auction increases in the bid:
the higher the bid, the higher the probability of being displayed. When the budget smoothing
is not initiated, then $\pi_i=1$ and the probability of appearing on the page is simply $eQ_i(b_i)$
(increasing in the bid). When the bid exceeds the optimal level, then the probability of being eligible for an impression is
$
\pi_i=\mbox{Budget}_i /eCPM_i(b_i)$ leading to the probability of being allocated an impression
of
$$
\mbox{Budget}_i \times \frac{eQ_i(b_i)}{eCPM_i(b_i)}.
$$
This function decreases as a function of bid. This means that if the per impression budget
does not warrant a given bidder the top position without filtering, then the probability of getting an impression
increases up to the optimal bid level and then decreases whenever the bid starts exceeding the optimal level.

\paragraph{\textbf{Budget and bid recommendations based on the impression targets}}
We can use the ``expected impression" model to make the recommendations
for the choice of the monthly budget and the corresponding bid that meet
a given impression target. Note that due to the budget smoothing, the
expected spent in a given impression
$$
\mbox{Spent}_i(b_i)=\pi_i\,eCPM_i(b_i) \leq \mbox{Budget}_i.
$$
The inequality may not be binding due to the possible jumps
in the $eCPM$ curve. We note that the expected probability
of appearing in the impression is
$$
\mbox{Prob}_i\left(b_i,\,\mbox{Budget}_i\right)=\pi_i\,eQ_i(b_i).
$$
Consider this probability as a function of the bid and the budget,
taking into account our model of filtering due to budget smoothing.
$$
\mbox{Prob}_i\left(b_i,\,\mbox{Budget}_i\right)=
\left\{
\begin{array}{ll}
eQ_i(b_i),&\;\;\mbox{if}\; eCPM_i(b_i) \leq \mbox{Budget}_i,\\
\mbox{Budget}_i\frac{eQ_i(b_i)}{eCPM_i(b_i) },&\;\;\mbox{if}\; eCPM_i(b_i) > \mbox{Budget}_i.\\
\end{array}
\right.
$$
The expected impression count is obtained by multiplying the probability
of appearing on the page $\mbox{Prob}_i\left(b_i,\,\mbox{Budget}_i\right)$
by the total projected impression inventory.
Note that function $\mbox{Prob}_i\left(b_i,\,\mbox{Budget}_i\right)$
is increasing in the bid $b_i$ up to the bid $b_i^*(\mbox{Budget}_i)$
such that
$$
eCPM_i\left(b_i^*(\mbox{Budget}_i)\right) = \mbox{Budget}_i
$$
and decreases when the bid is greater than $b_i^*(\mbox{Budget}_i)$.
Therefore, the expected number of impressions is maximized
for a given budget at $b_i=b_i^*(\mbox{Budget}_i)$.

Let $\mbox{Inventory}$ be the total impression inventory and $\mbox{Goal}_i$
be the impression target for bidder $i$. Then the optimum bid for a given
budget is set as
$$
\mbox{Prob}_i\left(b_i,\,\mbox{Budget}_i\right) \leq\frac{\mbox{Goal}_i}{\mbox{Inventory}}.
$$
The minimum budget per impression for which the impression goal is met is
$$
\mbox{Budget}_i=eCPM_i(b_i),
$$
leading to
$$
eQ_i(b_i)=\frac{\mbox{Goal}_i}{\mbox{Inventory}}.
$$
To determine the recommendations of the bid and the budget
based on the expressions above, we formulate the following problem:
find the profile of the probabilities of eligibility for an auction $\pi_1,\pi_2,\ldots,\pi_I$
and the optimal bid of bidder $i$ $b_i^*$ such that
\begin{enumerate}
\item $\pi_i=1$ for bidder of interest $i$;
\item $eQ_i(b_i^*)=\frac{\mbox{Goal}_i}{\mbox{Inventory}}$.
\end{enumerate}
Note that this is equivalent to solving a system of equations
\begin{equation}\label{one bidder}
\begin{array}{ll}
\pi_j=\min\left\{\frac{\mbox{Budget}_j}{eCPM_j(b_j)},\,1 \right\},\,j \neq i,\\
\pi_i=1,\\
eQ_i(b_i^*)=\frac{\mbox{Goal}_i}{\mbox{Inventory}}.\\
\end{array}
\end{equation}
with unknowns $\pi_j$, $j \neq i$ and $b_i^*$. The recommended
bid is the solution $b_i^*$ and the budget recommendation is given
by multiplying the per impression budget
$$
\mbox{Budget}^*_i=eCPM_i(b_i^*)
$$
by the projected impression inventory.

We also need to account for possible corner solutions
that do not allow the equality $eQ_i(b_i^*)=\frac{\mbox{Goal}_i}{\mbox{Inventory}}$
to be satisfied. First, suppose that for the grid of bids $\{b_k^g=\frac{s_k\,b_k}{s_i}\}^I_{k=1}$
we observe
$$
\max_k\,eQ_i(b^g_k)<\frac{\mbox{Goal}_i}{\mbox{Inventory}}.
$$
Then the optimal bid $b_i^*=\max_k\,b^g_k$.
In that case we set the budget $\mbox{Budget}^*_i=b_i^*>eCPM_i(b_i^*)$.
The rationale for this is that the top bidder does not have an incentive to
set the bid below the budget as that would lead to a weak decrease in the
expected number of impressions while not decreasing the expected
spent.

Second, suppose that
$$
\min_k\,eQ_i(b^g_k)>\frac{\mbox{Goal}_i}{\mbox{Inventory}}.
$$
In that case for any bid level the bidder will be subject to budget
smoothing. Thus the optimal bid will correspond to
$$
b_i^*=\min_k\,b^g_k.
$$
The recommended budget will correspond to
$$
\mbox{Budget}_i^*=\frac{\mbox{Goal}_i}{\mbox{Inventory}}\,
\frac{eCPM_i(b_i^*) }{eQ_i(b_i^*)}.
$$

\paragraph{\textbf{Simultaneous optimization for multiple bidders}}
The automated optimization for multiple bidders is based on
a simple generalization of the single bidder problem.
We note that for all bidders whose bids and budgets are optimized
to meet the impression goals, we need to solve
a system of equations equivalent to (\ref{one bidder}) to find
bids $b_j^*$. Let $\mathcal J$ be the subset of bidders
who use the bid and budget recommendation. Then we
find the set of recommended bids $\{b_j^*,\,j \in {\mathcal J}\}$
by solving the system of equations

\begin{equation}\label{many bidders}
\begin{array}{l}
\pi_k=\min\left\{\frac{eCPM_k(b_k)}{\mbox{Budget}_k},\,1 \right\},\,k \not\in {\mathcal J},\\
\pi_k = 1,\,k \in {\mathcal J}\\
eQ_k(b_k^*)=\frac{\mbox{Goal}_k}{\mbox{Inventory}},\,k \in {\mathcal J}.\\
\end{array}
\end{equation}

We also take into account the ``corner solutions" corresponding to very low
and very high impression goals relative to the available inventory.

\paragraph{\textbf{Formal integrity tests for tool performance}}
The structure of the rank-based auction leads to the set of properties that
have to be satisfied by the optimal solutions for bids and budgets.
We can use these properties to construct the tests for the performance
of the recommendation tool.
\begin{enumerate}
\item For any $\tau>0$, if $b_k,\,k=1,\ldots,I$ is the solution of (\ref{many bidders}),
then if the filtering probabilities are fixed, then the replacement
of $b_k$ with $\tau\, b_k$ does not change the predicted impression counts
$eQ_k(\tau\,b_k)\times \mbox{Inventory}$.
\item For any $\tau>0$, if $b_k,\,k=1,\ldots,I$ is the solution of (\ref{many bidders}),
then if the filtering probabilities are fixed, then the replacement
of $b_k$ with $\tau\, b_k$ leads to the proportional increase in the predicted total spent
$eCPM_k(\tau\,b_k)\times \mbox{Inventory}=\tau\,eCPM_k(b_k)\times \mbox{Inventory}$.
\item For any $\tau>0$, if the inventory changes to $\tau\,\mbox{Inventory}$
and all impression goals change to $\tau\,\mbox{Goal}_i$, then the optimal
bids and per impression recommended budgets remain the same.
\item The ratio $\frac{eQ_i(b_i)}{eCPM_i(b_i)}$ is a (weakly) monotone decreasing
function of the bid. In other words, for grid points $\{b_k^g\}^I_{k=1}$,
$b^g_m>b^g_n$ should lead to $\frac{eQ_i(b_m^g)}{eCPM_i(b_m^g)}<\frac{eQ_i(b_n^g)}{eCPM_i(b_n^g)}$.
\end{enumerate}

\section{Methodology for estimation of values and regret.}
Recall the notion of regret and rationalizable set from Section \ref{sec:trust}.
The structure of the rationalizable set for 9 of the bidders most frequently changing bids in each of the 6 markets we analyzed is shown in Figures \ref{Fig:HistBidChangeFrequencies:12}-\ref{Fig:HistBidChangeFrequencies:56}.
The rationality assumption of the inequality (\ref{eqn:eps-regret}) models
players who may be learning from experience while participating in
the game. We assume that the strategies $b_{it}$ and environment
parameters $\theta^t$ are input simultaneously, so agent $i$
cannot pick his strategy dependent on the state of nature $\theta^t$
or the strategies of other agents $b_{-i,t}$. This makes the
standard of a single best strategy $b$ natural, as chosen
strategies cannot depend on $\theta^t$ or $b_{-i,t}$.
Beyond this, we do not make any assumption on what information
is available for the agents, and how they choose their strategies.

We can specialize the definition of the rationalizable set in (\ref{eqn:eps-regret})
to auctions for randomly arriving impressions by introducing functions
\begin{equation}
\Delta \mbox{eCPM}_i(b')= \frac{1}{T} \sum_{t=1}^{T}\left( \mbox{eCPM}_{it}(b')-\mbox{eCPM}_{it}(b_{it})\right),\;
\mbox{and}\;
\Delta eQ_i(b')=\frac{1}{T} \sum_{t=1}^{T} \left(eQ_{it}(b')-eQ_{it}(b_{it})\right),
\end{equation}
corresponding to an aggregate outcome in $T$ time periods from switching
to a fixed bid $b'$ from the actually applied bid sequence $\{b_{it}\}^T_{t=1}$.
The $\epsilon$-regret condition reduces to:
\begin{equation}\label{eqn:halfplanes}
\forall b'\in \R_+: v_i\cdot \Delta eQ_i(b') \leq \Delta \mbox{eCPM}_i(b') + \epsilon_i
\end{equation}
for each bidder $i$.
Hence, the rationalizable set $\NR$ is an
envelope of the family of half planes obtained by varying $b \in {\mathbb R}_+$ in Equation \eqref{eqn:halfplanes}.

Under suitable assumptions regarding the expected auction outcomes
$eQ_{it}(\cdot)$ and $eCMP_{it}(\cdot)$ in bidder $i$'s bid, such as continuity and
monotonicity, one can
establish basic geometric properties of the rationalizable set,
such as its convexity and closedness. \cite{NST:2015} find a simple
geometric characterization of the $\NR$ set that also implies an efficient
algorithm for computing that set. Since closed convex bounded sets
are fully characterized by their boundaries, we can use
the notion of the {\it support} function to represent the
boundary of the set $\NR$. The support function of a closed convex set $X$ is
$
h(X,u)=\sup_{x \in X}\langle x,u\rangle,
$
where in our case $X=\NR$ is a subset of ${\mathbb R}^2$ or value and error pairs $(v_i,\epsilon_i)$, and  then $u$ is also an element of ${\mathbb R}^2$.

An important property of the support function
is the way it characterizes closed convex bounded sets. Denote
by $d_H(A,B)$ the Hausdorf distance between convex compact sets $A$ and $B$.
Recall that the Hausdorf norm for subsets
$A$ and $B$ of the metric space $E$ with metric $\rho(\cdot,\cdot)$ is defined
as
$$
d_H(A,B)=\max\{\sup\limits_{a \in A}\inf\limits_{b \in B}\rho(a,b),\,
\sup\limits_{b \in B}\inf\limits_{a \in A}\rho(a,b)\}.
$$
It turns out that $d_H(A,B)=\sup_u|h(A,u)-h(B,u)|$.
Therefore, if we find $h(\NR,u)$, this will be equivalent
to characterizing $\NR$ itself. The following result
fully characterizes the support function of the set $\NR$
based on the aggregate auction ouctomes $\Delta \mbox{eCPM}_i(\cdot)$
and $\Delta eQ_i(\cdot)$:
\begin{theorem}
Under monotonicity of $\Delta \mbox{eCPM}_i(\cdot)$
and $\Delta eQ_i(\cdot)$ the support function
of $\NR$ is function $h\,:\,\{(u_1,u_2)\,:\,u_1,u_2 \in \R,\;u_1^2+u_2^2=1\} \mapsto \R_+$ such that
$$
h(\NR,u)=\left\{\begin{array}{ll}
|u_2| \Delta eQ_i\left(\Delta \mbox{eCPM}^{-1}_i\left(\frac{u_1}{|u_2|}\right)\right),\,\mbox{if}\,
u_2<0,&\; \mbox{if}\,\frac{u_1}{|u_2|}\in \left[\inf_b \Delta \mbox{eCPM}_i,\,
\sup_{b}\Delta \mbox{eCPM}_i(b)\right]\;\;
\\
+\infty, & \mbox{otherwise}.
\end{array}
\right.
$$
\end{theorem}

This theorem is the identification result for valuations and algorithm parameters
for $\epsilon$-regret learning algorithms. Unlike equilibrium settings that we discussed above,
we cannot pin-point the values of players.
At the same time, the characterization of the set $\NR$ reduces to
evaluation of two one-dimensional functions. We can use efficient
numerical approximation for such an evaluation. The shape
of the set $\NR$ will generally depend on the parameters of a concrete
algorithm used for learning. Thus the analysis of the geometry of $\NR$
can help us not only to estimate valuations of players but the learning algorithm as well.

The inference for the set $\NR$ reduces to the characterization of its
support functions which only requires to evaluate the function
$\Delta e Q_i\left(\Delta \mbox{eCPM}_i^{-1}\left(\cdot\right)\right)$. It is a one-dimensional
function and can be estimated from the data via direct simulation.

Since our object of interest is the set $\NR$ we need to characterize the
distance between the true set $\NR$ and the set $\widehat{\NR}$
that is obtained from subsampling the data. \cite{NST:2015} show
that the characterization of the properties of the estimated rationalizable set
reduces to the description of the properties of a single dimensional function
$
f(\cdot)=\Delta eQ_i\left(\Delta \mbox{eCPM}_i^{-1}\left(\cdot\right)\right)
$.
and let $\widehat{f}(\cdot)$ be its empirical analog recovered from the data.
The set $\NR$ is characterized by its support function
$h(\NR,u)$.
Then using the relationship between the Hausedorf norm
and the sup-norm of the support functions we can write
$$
d_H(\widehat{\NR},\,\NR)=\sup\limits_{\|u\|=1}|h(\widehat{\NR},u)-h(\NR,u)|
\leq \sup\limits_{z}\left|
\widehat{f}\left(z\right)-
{f}\left(z\right)\right|,
$$
The empirical analog of function  $f(\cdot)$ can be directly estimated
from the data via subsampling of auctions.
The properties of the estimated set $\widehat{\NR}$ are thus determined
by the properties of function $f(\cdot)$. In particular, if
function $f$ has derivative up to order $k \geq 0$ and for some $L \geq 0$,
$
|f^{(k)}(z_1)-f^{(k)}(z_2)| \leq L|z_1-z_2|^{\alpha},
$
then with probability approaching 1 the Hausedorf distance between the true and estimated
rationalizable set can be bounded as
$$
d_H(\widehat{\NR},\,\NR) \leq O((N^{-1}\log\,N)^{\gamma/(2\gamma+1)}),\;\;\gamma=k+\alpha
$$
with probability approaching 1 as $N \rightarrow \infty$,
where $N=n \times T$ is the total number of samples available (with $T$ auctions and $n$ players
in each).

\section{Additional tables and figures.}
\begin{figure}[!ht]
\begin{center}
  \includegraphics[width=6.5in]{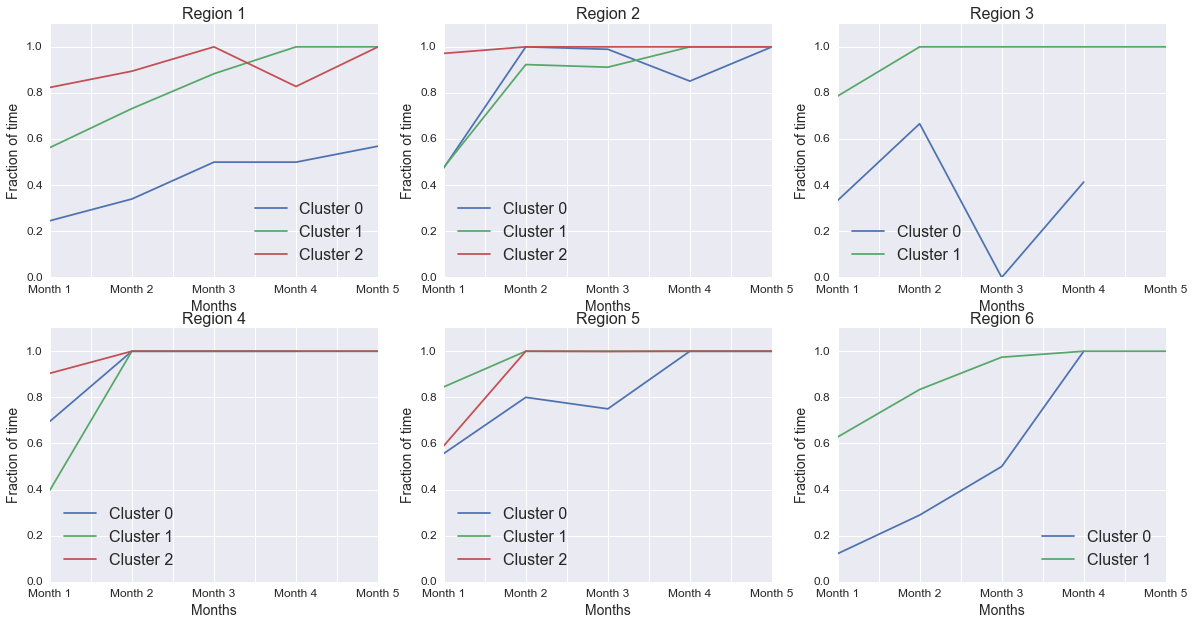}
  \end{center}
 \caption{Average fraction of time agents follow the recommended bid separated by clusters and regions.}
  \label{Fig:FollowingSugBidRegions}
 \end{figure}

\begin{figure}[!th]
\begin{center}
  \includegraphics[width=6.5in]{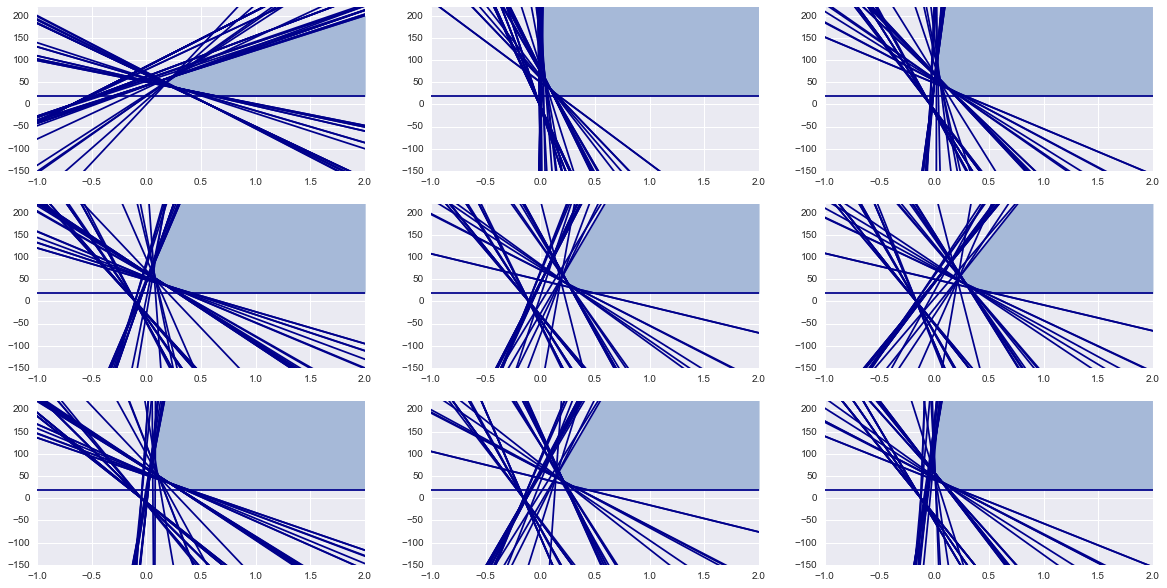}\\
  {\small Region 1}\\
   \includegraphics[width=6.5in]{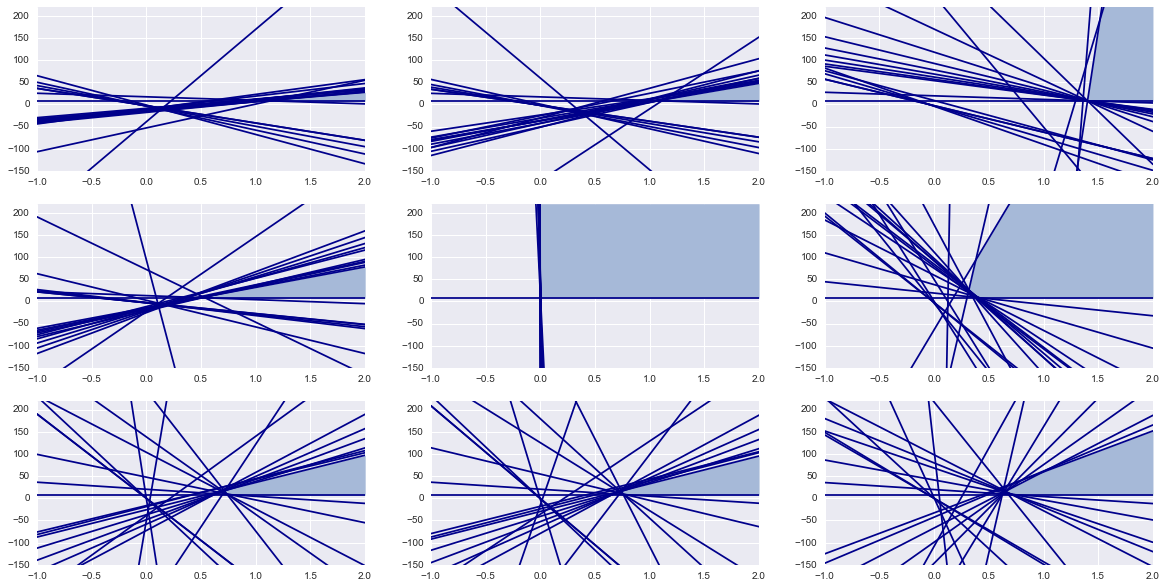}\\
   {\small Region 2}
  \caption{Rationalizable set for 9 agents most frequently changing bids}
  \label{Fig:HistBidChangeFrequencies:12}
  \end{center}
\end{figure}
\newpage
\begin{figure}[!th]
\begin{center}
  \includegraphics[width=6.5in]{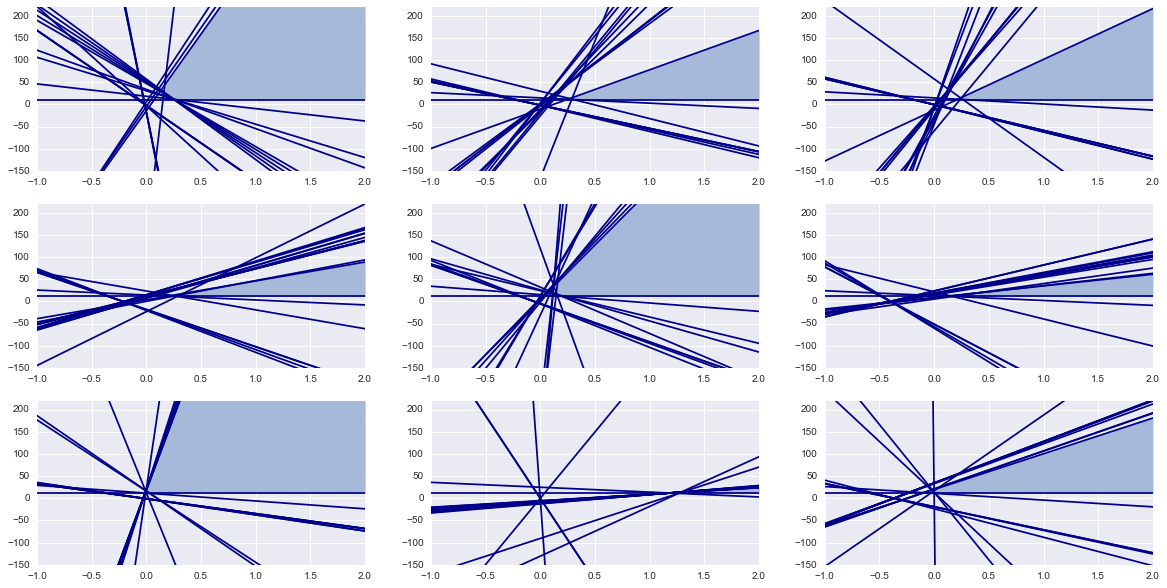}\\
  {\small Region 3}\\
   \includegraphics[width=6.5in]{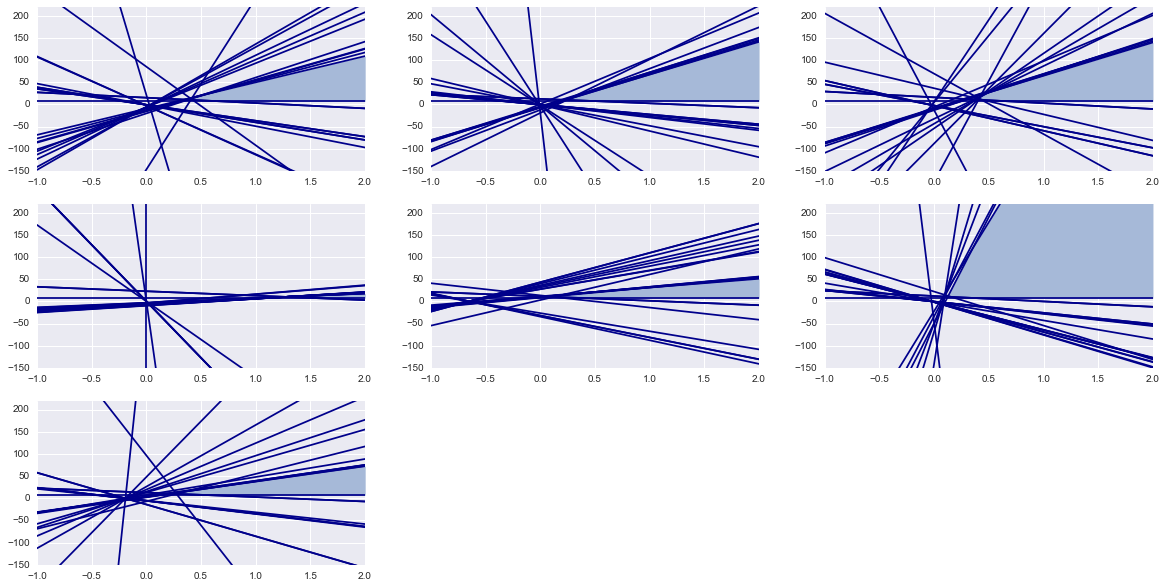}\\
   {\small Region 4}
  \caption{Rationalizable set for 9 agents most frequently changing bids}
  \label{Fig:HistBidChangeFrequencies:34}
  \end{center}
\end{figure}
\newpage
\begin{figure}[!th]
\begin{center}
  \includegraphics[width=6.5in]{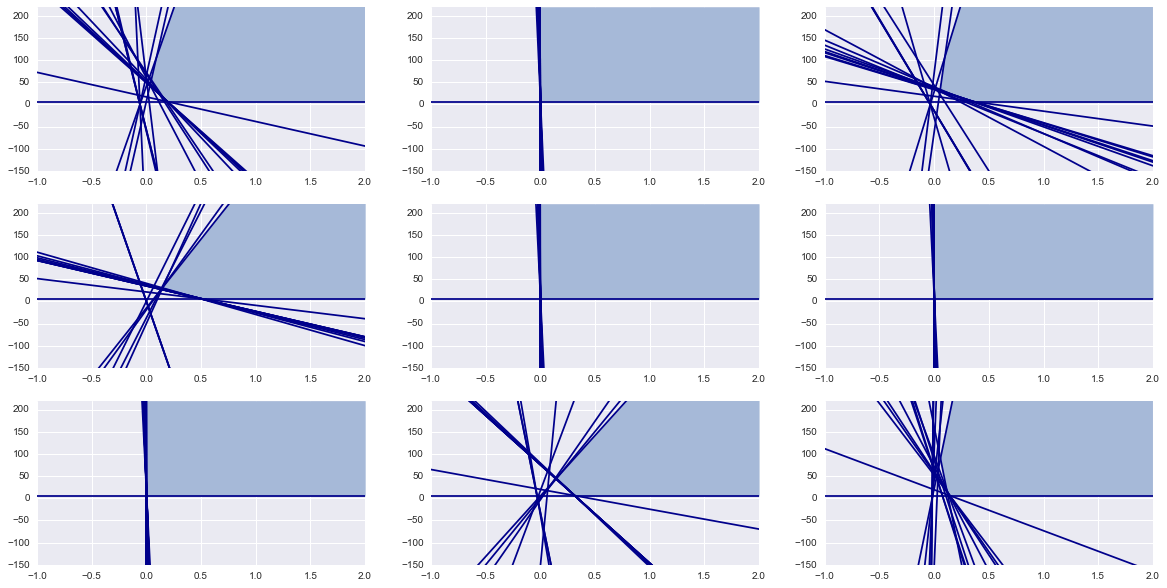}\\
  {\small Region 5}\\
   \includegraphics[width=6.5in]{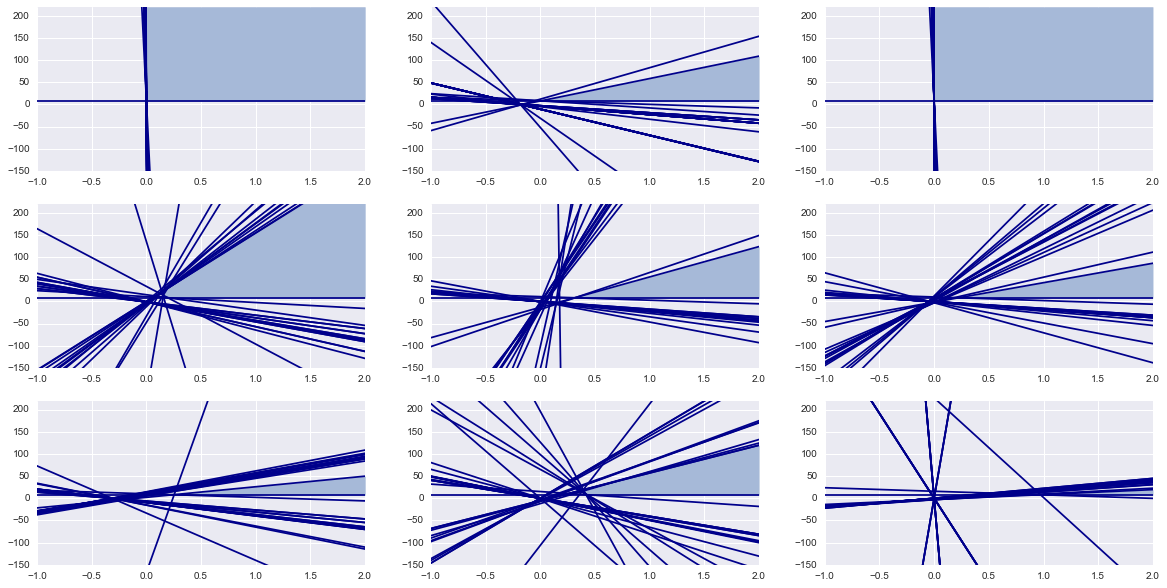}\\
   {\small Region 6}
  \caption{Rationalizable set for 9 agents most frequently changing bids}
  \label{Fig:HistBidChangeFrequencies:56}
  \end{center}
\end{figure}

\end{document}